\documentclass{aa}

\usepackage{graphicx}
\usepackage{dcolumn}
\newcolumntype{d}[1]{D{±}{\pm}{#1}}

\usepackage{txfonts}
\usepackage{xcolor}
\usepackage{hyperref}
\hypersetup{
  colorlinks=true,
    linkcolor=blue,
    citecolor=blue,
    urlcolor=blue}

\usepackage{amsmath}
\usepackage{pgf,tikz,pgfplots}
\pgfplotsset{compat=1.15}
\usepackage{mathrsfs}
\usepackage{physics}

\usepackage{siunitx}
\usepackage{subcaption}
\usepackage[switch]{lineno} 
\usepackage{caption}

\def\fdnu{f_{\Delta \nu}}
\def\msun{{\rm M}_{\odot}}
\def\rsun{{\rm R}_{\odot}}
\def\fnumax{f_{\numax}}
\newcommand{\alphafe}{[\alpha/\mathrm{Fe}]}

\newcommand{\nuny}{\nu_{\mathrm{Nyq}}}
\newcommand{\numax}{\nu_{\mathrm{max}}}
\newcommand{\nusub}{\nu_{\mathrm{max,sub}}}
\newcommand{\dnu}{\Delta\nu}
\newcommand{\kep}{\textit{Kepler }}
\newcommand{\teff}{T_\mathrm {eff}}
\newcommand{\logg}{\log g}
\newcommand{\afe}{[$\alpha$/Fe]}
\newcommand{\feh}{[Fe/H]}
\newcommand{\gaia}{Gaia}
\newcommand{\teffsun}{{\rm T}_{\mathrm {eff},\odot}}
\DeclareMathOperator{\sinc}{sinc}
\usepackage[OT2,T1]{fontenc}
\DeclareSymbolFont{cyrletters}{OT2}{wncyr}{m}{n}
\DeclareMathSymbol{\Sha}{\mathalpha}{cyrletters}{"58}

\usepackage{siunitx} 
\begin{document}

   \title{Beyond the Nyquist frequency}

   \subtitle{Asteroseismic catalog of undersampled \textit{Kepler} late~subgiants~and~early~red~giants}
    \author{B. Liagre
          \inst{1,2,3,4}
          \and
          R.~A. Garc\'ia\inst{2}
          \and 
          S. Mathur \inst{3,5}
          \and
          M.~H. Pinsonneault \inst{6}
          \and
          A. Serenelli\inst{7,8}
          \and
          J.~C. Zinn\inst{9}
          \and
          K.~Cao \inst{11}
          \and
          D.~Godoy-Rivera\inst{3,5}
          \and
          J. Tayar\inst{10}
          \and          
          P.~G.~Beck \inst{3,5}
          \and
          D.~H. Grossmann\inst{3,5}
          \and
          D.B. Palakkatharappil \inst{2}
          }

   \institute{
            ENS Paris-Saclay, Université Paris-Saclay, 91190, Gif-sur-Yvette, France
           \and
            Universit\'e Paris-Saclay, Universit\'e Paris Cit\'e, CEA, CNRS, AIM, F-91191, Gif-sur-Yvette, France
         \and
            Instituto de Astrof\'isica de Canarias (IAC), E-38205 La Laguna, Tenerife, Spain
          \and 
          Institute of Science and Technology Austria (ISTA), Am Campus 1, 3400 Klosterneuburg, Austria
          \and
           Universidad de La Laguna (ULL), Departamento de Astrof\'isica, E-38206 La Laguna, Tenerife, Spain
           \and
            Department of Astronomy, The Ohio State University, Columbus, OH 43210, USA
            \and 
            Institute of Space Sciences (ICE, CSIC), Carrer de Can Magrans S/N, E-08193, Cerdanyola del Vallès, Spain
         \and
             Institut d’Estudis Espacials de Catalunya (IEEC), Carrer Esteve Terradas, 1, Edifici RDIT, Campus PMT-UPC, E-08860, Castelldefels, Spain
        \and
            Department of Physics and Astronomy, California State University, Long Beach, Long Beach, CA 90840, USA
        \and
            Department of Astronomy, University of Florida,  Gainesville, FL 32611, USA   
           \and
            Department of Physics, The Ohio State University, Columbus, OH 43210, USA
            }

   \date{Received: 2025; accepted: 24/07/2025 }

  \abstract
    {Subgiants and early red giants are crucial for studying the first dredge-up, a key evolutionary phase where the convective envelope deepens, mixing previously interior-processed material and bringing it to the surface. 
    Yet, very few have been seismically characterized with \textit{Kepler} because their oscillation frequencies are close to the 30 minute sampling frequency of the mission.
    We developed a new method as part of the new PyA2Z code to identify super-Nyquist oscillators and infer their global seismic parameters, $\nu_\mathrm{max}$ and large separation, $\Delta\nu$. 

    Applying PyA2Z to 2\,065 \emph{Kepler} targets, we seismically characterize 285 super-Nyquist and 168 close-to-Nyquist stars with masses from 0.8 to 1.6 $\msun$. In combination with APOGEE spectroscopy, Gaia spectro-photometry, and stellar models, we derive stellar ages for the sample. There is good agreement between the predicted and actual positions of stars on the HR diagram (luminosity vs. effective temperature) as a function of mass and composition. While the timing of dredge-up is consistent with predictions, the magnitude and mass dependence show discrepancies with models, possibly due to uncertainties in model physics or calibration issues in observed abundance scales. 
   }


   \maketitle

\section{Introduction}

Turbulence in the outer layers of cool stars excites oscillations
 \citep{1977ApJ...212..243G,1988ApJ...328..879K}.
 Asteroseismology, the characterisation of the properties of stars through the analysis of oscillation modes \citep{1984srps.conf...11C}, has proven to be a valuable tool in studying solar-like pulsating stars. The characteristic oscillation frequencies are proportional to the mean density; they range from 5 minutes for solar analogues, hours to days for low- and mid- luminous red giants stars (RG), of order 5 hours for core He-burning (RC) stars, to months or years for the most luminous ones on the red giant branch (RGB) \citep{2019A&A...622A..76M}.
 
 With the advent of large time-domain surveys from space missions, these oscillations became detectable for large numbers of stars. However, previous studies have generally been dominated by evolved RGs, which have higher amplitudes and lower oscillation frequencies. These frequencies can generally be detected in the baseline observing mode of the instruments performing the surveys e.g. Long Cadence (LC) in the \emph{Kepler} mission \citep{2010Sci...327..977B}, Full Frame image data from the Transit Exoplanet Survey Satellite \citep[TESS,][]{2014SPIE.9143E..20R} and therefore require less specific and deliberate targeting. Several hundred thousand of RGs on the RGB and Red Clump (RC) were analysed using asteroseismology \citep[e.g.][]{2009Natur.459..398D,2010ApJ...723.1607H,2010A&A...517A..22M,2015ApJ...809L...3S,2017ApJ...835...83S,2016A&A...597A..30A,2018ApJS..236...42Y,2020ApJS..251...23Z,2022ApJ...926..191Z,2021ApJ...919..131H,2022Natur.610...43L}. 
 
 Seismic observers have also been particularly interested in the solar-like oscillations of main-sequence dwarf stars \citep{2019LRSP...16....4G}. They require special observation modes with shorter cadences. By contrast, only around 1000 main sequence solar-like oscillators have been seismically characterized \citep[e.g.][]{2009A&A...507L..13B,2011A&A...530A..97B,2015MNRAS.446.2959D,2016MNRAS.456.2183D,2017ApJS..233...23S,
2017ApJ...835..173S,2021ApJ...922...18O,2022AJ....163...79H,2022A&A...657A..31M,2023A&A...674A.106G,2023A&A...669A..67H,2024A&A...688A..13L}. 
 The domain between the main sequence and the giant branch, referred to as the subgiant branch, is astrophysically interesting. However, stars in this domain were neither targeted as planet-host candidates nor exhibited detectable oscillation modes in long-cadence data. As a consequence, this subgiant transition zone is much more poorly understood, and it is clear that important physical processes occur in cool subgiants. At this evolutionary stage, nuclear burning ceases in the core and fusion is ignited in a shell surrounding it. The core contracts and the envelope expands in these stars, affecting their radial rotation profile that exhibits strong internal differential rotation  \ 
\citep[e.g.][]{2012Natur.481...55B,2012ApJ...756...19D,2014A&A...564A..27D}. This core contraction also influences their magnetic fields and activity \citep[e.g.][]{2014A&A...563A..84G,2021ApJ...915...19G,2024ApJ...974...31M}. The hydrogen-exhausted cores also become dense, and at low mass degeneracy pressure sets in. At the same time, the fully mixed surface convection zone deepens. This causes the products of core nuclear processing appear at the surface as stars develop deep surface convective envelopes \citep{2024MNRAS.530..149R}. This first dredge-up \citep{1965ApJ...142.1447I,2015A&A...583A..87S} has been used as a mass and age diagnostic for stellar populations \citep{Martig2016, Ness2016} and is a key prediction of stellar theory. All of these processes are very sensitive to the details of the physics of the stellar interior and impact the future evolution of the star, including the expected nucleosynthesis \citep{KarakasLattanzio2014} and the resulting remnant \citep{Hermes2017}.
Unfortunately, only a small handful of subgiant and lower giant stars have the asteroseismic observations necessary to constrain their masses and probe their interior structures.

This relative lack of subgiant data is largely due to the observing strategies employed by space missions. The first generation of space time-domain missions, such as Convection Rotation and planetary Transit \citep[CoRoT,][]{2006cosp...36.3749B}, \textit{Kepler}, and K2 \citep{2014PASP..126..398H}, use relatively long observing baseline cadences optimized for transits of Earth-like planets around Sun-like stars.
For example, \emph{Kepler}  offers two acquisition modes: short cadence \citep[SC, sampling period $T_\mathrm{samp} = 58.85$~s and Nyquist frequency (defined as $\nuny=1/2T_\mathrm{samp}$) $\nuny \simeq 8\,496$ µHz,][]{2010ApJ...713L.160G} and long cadence \citep[LC, sampling period $T_\mathrm{samp}=29.4$~min and $\nuny \simeq 283$ µHz,][]{2010ApJ...713L.120J}. While most of the stars were observed in LC mode \citep[$197\,096$ targets for \textit{Kepler},][]{2017ApJS..229...30M,Diego_2025}, at any time, only 512 slots were allocated for SC observations. Although subgiant stars could be observed in this SC mode, the asteroseismic community privileged the study of main-sequence stars and early subgiants. As a result, only a few late subgiants and early RGBs were targeted.

\cite{2013MNRAS.430.2986M} showed that it was possible to recover the parameters of $\delta$ Scuti oscillating above the Nyquist frequency and \cite{2014MNRAS.445..946C} showed that cool, evolved subgiants and stars lying at the base of the red-giant branch, whose oscillation frequencies are well above the $\nuny$, could be studied with the LC \emph{Kepler} dataset by using the aliased frequencies below $\nuny$. They called it ``Super-Nyquist asteroseismology''. This methodology was further exploited by \cite{2016ApJ...827...50M} and \citet{2016MNRAS.463.1297Y} providing detection of oscillations in 47 stars in the super-Nyquist regime, up to a frequency of maximum power, $\numax\,= 387$ µHz, using LC \emph{Kepler} data.

In this paper, we develop a new methodology to study the pattern of modes around $\nuny$ and disentangle whether those modes are sub-Nyquist, super-Nyquist, or if they are centred around Nyquist (hereafter close-to-Nyquist). This results in a large sample of late subgiants and early red giants that allow us to study the first dredge up. Our work complements the recent APOKASC-3 paper \citep[hereafter APOKASC-3,][]{apokasc3}, which provided asteroseismic data for stars with {\it Kepler} light curves and Apache Point Observatory Galactic Evolution Experiment 
\citep[APOGEE,][]{2017AJ....154...94M} spectra. In that study, it was found that different methods diverged for $\numax > 220$\,$\mu$Hz, so no stellar parameters were inferred for such stars there. The upcoming García et al. catalog (in prep.) provides similar data for stars with $\numax < 220$\,$\mu$Hz, that do not have APOGEE spectra.  In this work, our analysis includes all {\it Kepler} stars with $\numax > 200$\,$\mu$Hz, and complements these other two papers.
In section~\ref{sec:Data preselection}, the observations used in this paper as well as the sample selection of stars are described. The principles behind the classification of stars as sub-Nyquist, close-to-Nyquist, or super-Nyquist are described in section~\ref{sec:sn detection} and the results are presented and discussed. 
In section~\ref{sec:MRA}, we compute the stellar parameters (masses, radii, and ages) from stellar models and we discuss how the sample probes the first dredge up phase in section~\ref{sec:discussion}. The conclusions are given in section~\ref{sect:conclusion}.

\section{Sample selection and data preparation}
\label{sec:Data preselection}
One of the objectives of this paper is to compile the most complete catalog of pulsating late subgiants and early red giants observed by \kep in long cadence, crucial for advancing in asteroseismology of solar-like stars with $\numax > 200$ µHz and study the dredge up. A paramount consideration in this endeavor is the avoidance of any selection bias, which holds immense significance for Galactic archaeology and stellar population studies. 

Briefly, the initial set of stars in APOKASC-3 and García et al. (in prep.) is a compilation of 30\,336 \emph{Kepler} targets selected to be potential subgiants and red giants.

The starting point is the catalog of 16\,094 confirmed RGs from \citet{2018ApJS..236...42Y}, which included previous detections  \citep{2011MNRAS.414.2594H,2011ApJ...743..143H,2013ApJ...765L..41S,2014ApJS..211....2H,2016ApJ...827...50M,2016MNRAS.463.1297Y}. Then, subgiants and RGs not included in the previous set were added from the \emph{Kepler} DR25 catalog \citep{2017ApJS..229...30M}, the \emph{Kepler}-Gaia DR2 catalogue \citep{2018ApJ...866...99B},  the DR16 and DR17 of the APOGEE spectroscopic survey \citep{2020ApJS..249....3A,2022ApJS..259...35A}, and other seismic RGs found during the visual inspections done for the surface rotation-period catalogues by \citet{2019ApJS..244...21S} and \citet{2021ApJS..255...17S}. 

From these sample, our stars are selected following a multi-step process. We first select stars from the APOKASC-3 catalog (i.e. with APOGEE spectra, see section~\ref{sec:spectra}) with a seismic $\nu_{\rm max}$ > 200 µHz. Then, we add stars from García et al. (in prep.), not included in APOKASC-3, and with a seismic $\nu_\mathrm{max}$ > 200 µHz. This differs from the $\nu_{\max} = 220$ µHz cutoff in the APOKASC-3 sample, as our goal was to ensure the most complete initial sample considering larger uncertainties on the García et al. (in prep.) values. This led to an initial sample of 1959 stars with seismic inferences. This sample will be referred to as the 'main sample' in the rest of the paper. Finally, we include 106 stars from the APOKASC-3 catalog without seismic detections (i.e. not already included in our sample) but with a predicted frequency of maximum power from spectroscopy, $\nu_\mathrm{max,spec}$ > 180 µHz \citep[as already done in e.g.][]{2018A&A...612A..22B}. This sample will henceforth be referred to as the spectroscopic sample. 

This threshold of 180 µHz is lower than the 200 µHz one used for the main sample to better account for the larger uncertainties on the spectroscopic parameters ($\sigma_{\logg} = 0.065$).
This predicted $\nu_\mathrm{max,spec}$ is computed from seismic scaling relations \citep{1991ApJ...371..396B,1995A&A...293...87K} as follows: 
\begin{equation}
\nu_\mathrm{max,spec}=\nu_{\mathrm{max,spec},\odot}\frac{g_\mathrm{spec}}{\mathrm{g}_{\odot}}\left(\frac{\teff}{\teffsun}\right)^{-1/2} \, ,
\end{equation}
where $g_\mathrm{spec}$ and $\teff$ are respectively the APOGEE spectroscopic surface gravity and effective temperature of the star, those parameters and the way we obtained them will be further developed in section \ref{sec:spectra}.

The \emph{Kepler} seismic analyses are done using KEPSEISMIC\footnote{\url{https://archive.stsci.edu/prepds/kepseismic/}} light curves filtered at 20 days obtained from the Mikulski Archive for Space Telescopes (MAST) archive. They are corrected with the \emph{Kepler} Asteroseimic data analysis and calibration Software \citep[\texttt{KADACS},][]{2011MNRAS.414L...6G},
to remove outliers, correct any jumps and drifts, and stitch together the \emph{Kepler} quarters. All the gaps are interpolated using in-painting techniques based on a multi-scale discrete cosine transform \citep{2014A&A...568A..10G,2015A&A...574A..18P}.

\section{Classification of close-to and super-Nyquist stars}
\label{sec:sn detection}

When the oscillations of a star are undersampled, its power spectrum undergoes a symmetry operation and its mode frequencies are reflected about the $\nuny$ frequency.

\begin{figure}[h!]
    \centering
    \includegraphics[scale=0.4]{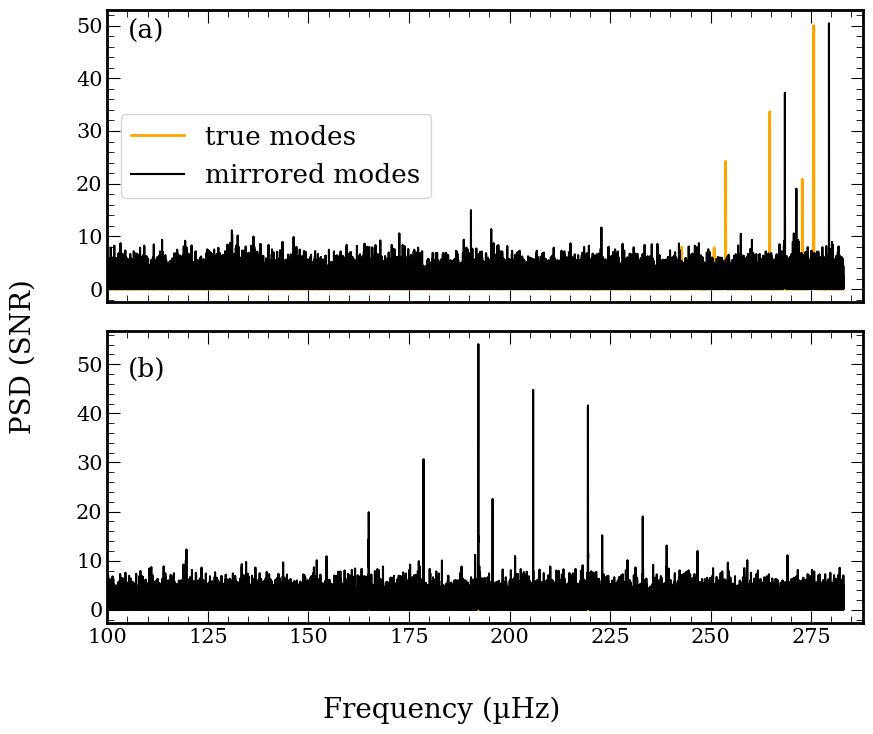}
    \caption{Power spectral density (PSD) of two simulated stars in units of signal-to-noise ratio (SNR) as a function of frequency. Panel (a): star with modes close to $\nuny$ at the right edge of the frequency range. The true modes are in orange and the mirrored modes in black. Panel (b): star with super-Nyquist modes where only the folded modes are observed between $0$ µHz and $\nuny$.}
    \label{fig:supnyvsny}
\end{figure}

This paper will consider two cases of undersampling: stars with modes very close to the Nyquist frequency (hereafter close-to-Nyquist stars) and stars with modes above the Nyquist frequency (super-Nyquist stars). For the close-to-Nyquist stars, when the modes above \(\nuny\) are reflected, they overlap with the ones below \(\nuny\) (see Fig. \ref{fig:supnyvsny}, panel (a)). For the super-Nyquist stars, when the modes are reflected below \(\nuny\) (see Fig. \ref{fig:supnyvsny}, panel (b)), they remain isolated under the Nyquist frequency.

Symmetry is isometric, meaning that folding preserves the distances between modes. In particular, for a star where the frequencies of the modes follow the asymptotic distribution \citep{tassoul1980asymptotic}, the frequencies of the modes $\nu_{n,l}$ roughly follow the pattern:

\begin{equation}\label{eq:Tassoul}
    \nu_{n,l} \approx \left(n+\frac{l}{2}+\varepsilon\right)\dnu\quad,
\end{equation}
where $n$ is the order of the mode, $l$ is its degree, and $\varepsilon$ is the phase term. This equation means that the modes of odd angular degree are found halfway of the modes of even degree. The dipole modes will be in between the radial modes hence creating a periodicity in $\frac{\dnu}{2}$ in the Power Spectrum Density (PSD). This periodicity is preserved by symmetry so long as the spectrum is not folded onto itself. 
Moreover, when getting closer to $\numax$, the PSD of the PSD will show an increase in power at $\frac{2}{\dnu}$. These phenomena allow the procedure to search for the seismic parameters to measure both $\numax$ and $\dnu$.

By isometry of the symmetry operation, the $\dnu$ measured on a folded spectrum will be correct. 
As for $\numax$, if the real modes were indeed below $\nuny$ then the measured $\numax$ ($\nu_{\mathrm{max,measured}}$) is the correct one. However, if the real modes were above the Nyquist frequency, we carried the measurement of $\numax$ on a symmetrized oscillation spectrum. Hence, in this case, the real $\numax$ should by symmetry be located at $2\nuny-\nu_{\mathrm{max,measured}}$.
\\ \par
Without prior information to find out whether a star has a folded spectrum or not, we
use a simple test: the pair of the set \{($\dnu$,$\nu_{\text{max,measured}}$),($\dnu$, $2\nuny-\nu_{\text{max,measured}}$)\} that best matches the seismic scaling relation $\dnu\propto\numax^{0.77}$ \citep{stello2009relation} will be assumed to be the correct pair of parameters.

\subsection{Measuring global seismic parameters in the sub- and super-Nyquist regimes}
\begin{figure}[h!]
    \centering
    \includegraphics[scale=0.4]{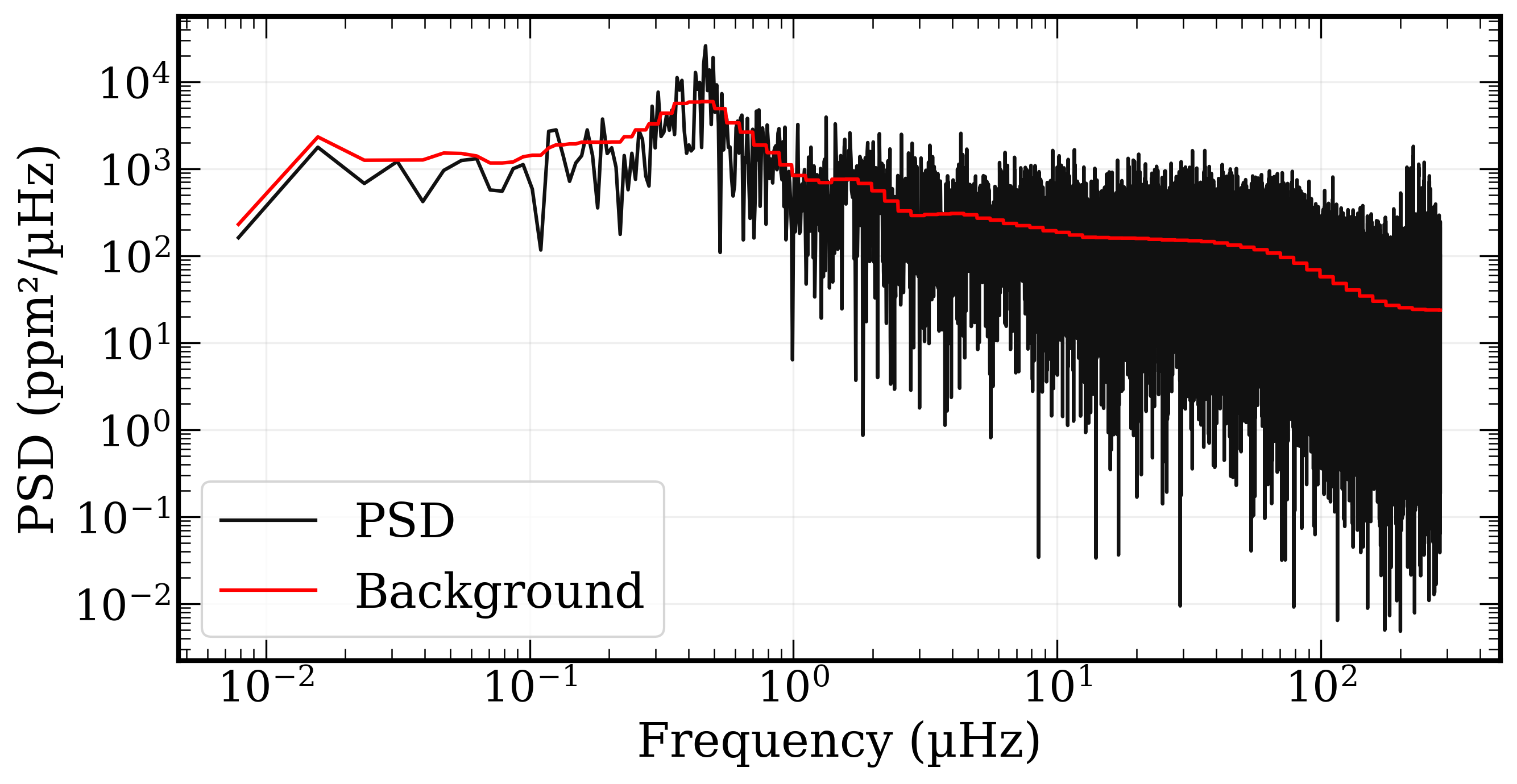}
    \caption{PSD of the star KIC 8179973 and estimate of its background with a median filter.}
    \label{fig:bkg}
\end{figure}

Before performing the seismic analysis, the components of the power spectrum arising from sources other than the oscillations, mostly convection and instrumental effects, should be removed. Generally, this is done by fitting the convective background with different components, including Harvey-like models \citep{1985ESASP.235..199H}, the p-mode envelope, and the photon noise \citep[e.g.][]{2011ApJ...741..119M,2014A&A...571A..71C,2014A&A...570A..41K}.
To flatten down the PSD, it is enough to use a median filter to characterize the background level at each frequency,  as shown in Fig. \ref{fig:bkg}. The resulting background estimate is then used to normalize the PSD by division.

\subsubsection{Measuring $\nusub$ of the sub-Nyquist PSD}\label{numax}
Without prior information on the star, it is not possible to discriminate between a sub- or super-Nyquist star. We apply the PyA2Z pipeline (Liagre et al., in prep.), which is the new Python version of the A2Z pipeline \citep{mathur2010determining}, to estimate the global seismic parameters of solar-like oscillations. 

We begin by determining the apparent $\numax$ below $\nuny$ (hereafter $\nusub$) without any prior guess. We note that in the case of a super Nyquist star, where the modes are mirrored across the Nyquist frequency, that this will not correspond to the true $\numax$ value. For instance in Fig. \ref{fig:supnyvsny} (b) the observed modes are those of a mirrored super Nyquist star and therefore, the directly measured $\nusub$ (201 µHz) will not correspond to the true $\numax$ (365 µHz).

\begin{figure}[h!]
    \centering
    \includegraphics[scale=0.36]{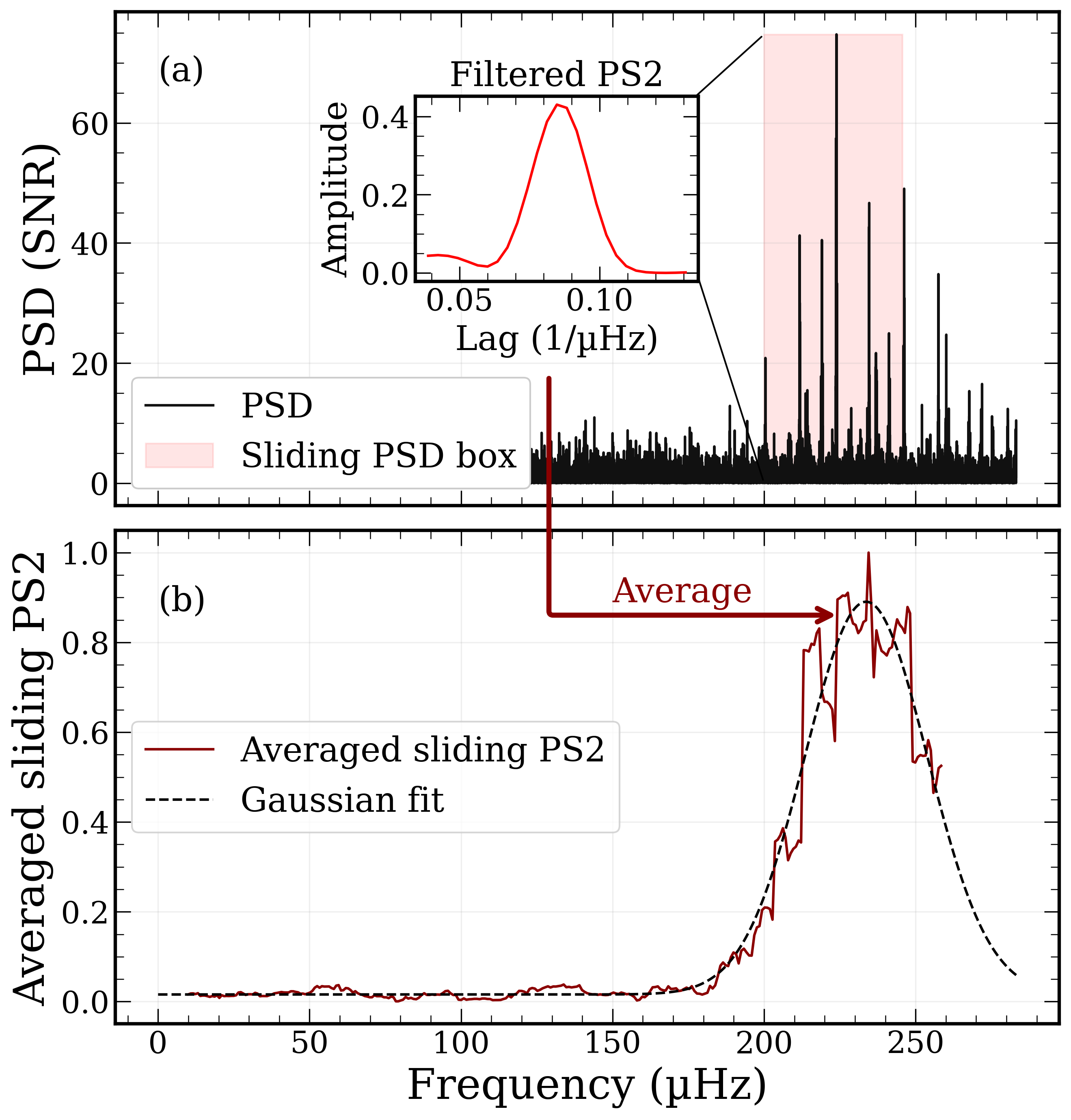}
    \caption{(a) PSD of the \emph{Kepler} star KIC 8179973 on which a sliding box (shaded pink area) is taken to compute the PS2 (inset) with $\Gamma=2$ and filtered as explained in Section \ref{numax}. (b) The red line is the averaged filtered PS2 normalized to its maximum value as a function of the center of the sliding windows. The dashed black line is the Gaussian fit.}
    \label{fig:nm_measure}
\end{figure}

To measure $\nusub$, PyA2Z relies on the periodicity of the acoustic modes in the PSD. 
As explained in Sec. \ref{sec:sn detection} the PSD of the PSD (hereafter PS2) in a region containing modes, should show a power excess around $\frac{2}{\dnu}$. The power excess should be at its maximum in the region around $\nusub$ assuming the mode amplitudes follow a Gaussian envelope.\\
Therefore, the pipeline measures $\nusub$ as follows (the process is illustrated on Fig. \ref{fig:nm_measure}):
\begin{enumerate}
    \item A segment of the PSD is selected on a box centered on $\nu_{\mathrm{center}}$ and of width $\Gamma\times\beta\times\nu_{\mathrm{center}}^{\alpha}$, we used a typical value of $\Gamma=2$. The parameters $\beta$ and $\alpha$ are given by the scaling relation $\frac{\dnu}{\text{µHz}}=\beta\left(\frac{\numax}{\text{µHz}}\right)^\alpha$ with $\alpha\approx 0.77$ following \cite{stello2009relation} and $\beta=\frac{\Delta\nu_{\sun}}{\nu_{\text{max},\sun}^{\alpha}}\text{µHz}^{\alpha-1}\approx 0.28$ being normalized to the solar value. The key idea behind this is that if $\nu_{\mathrm{center}}=\numax$ then we select~$\approx 3\times\Gamma$ peaks which is enough for computing the PS2 and detect the $\frac{\dnu}{2}$ periodicity without having to deal with unwanted noise. 
    If the box is too big, the $\chi^2$ noise of the PSD will pollute the PS2 and that can result in a reduced sensibility or a non-detection of the modes whereas if the box is too small, not enough modes will be present to ensure the periodicity detection by the PS2 method. Local deviations to the asymptotic pattern \citep[e.g. glitches,][]{2012AN....333.1040M,2014ApJ...782...18M} are not expected to have a significant impact on the measurement of $\numax$ as these would effectively be averaged out by our global analysis.
    \item The algorithm computes the PS2 of that segment and filters it around the expected positions of $\frac{1}{\dnu}$ and $\frac{2}{\dnu}$ (given by the scaling relation $\frac{\dnu}{\text{µHz}}=\beta\left(\frac{\numax}{\text{µHz}}\right)^\alpha$) assuming $\nu_{\text{center}}=\nusub$ (see inset of Fig. \ref{fig:nm_measure} (a)).
    \item The filtered PS2 is then averaged and that average is stored into an array $A$.
    \item The box is slid up to higher frequencies by a fraction $s$ of $\dnu$ and the algorithm goes back to step 1 until it reaches the Nyquist frequency.
    \item Once the box is at the Nyquist frequency, a Gaussian function is fitted to $A$ as a function of $\nu_{\text{center}}$.
    The mean of the fitted Gaussian function is then our measure of $\numax$. This measure also provides the standard deviation of the fitted Gaussian function $\sigma_{\text{nm}}$ which will be used as a proxy for the width of the amplitude envelope of the mode.
    \\
    Several tests with Markov Chain Monte Carlo, MCMC, methods have shown that very little uncertainty comes from the fitting process itself (typically less than $1$ µHz on the range of frequencies observed). \citet{2018ApJS..239...32P} showed that the uncertainty is largely dominated by the dispersion of the values obtained by different seismic pipelines. 
    For that reason, we ignore the errors coming from the fitting process and decide to conservatively assign an error bar of $\pm \frac{\dnu}{2}$ to our measurement for $\numax$. This error bar comes from the fact that, $\numax$ being an observational parameter, it is limited in precision by the density of modes in the envelope and these are asymptotically distinct by $\frac{\dnu}{2}$.\\

\end{enumerate}

\subsubsection{Measuring $\dnu$}\label{dnu}
Once $\numax$ is measured, the PSD is filtered by a zero padded hanning window around it. The filter $F$ is defined as follows:
\begin{equation}F(\nu)\left\{\begin{array}{cc}
     & 0 \text{ if } \left|\nu-\numax\right|>m\times\sigma_{\mathrm{nm}}  \\
     & \text{else }\frac{1}{2}\left(1 + 
                \cos\left(2\pi\frac{\nu-\numax}{m\sigma_{\mathrm{nm}}}\right)\right)
\end{array}\right.\quad,\end{equation}

where $m\approx 4.3$ has been found to give precise $\dnu$ without too much sensitivity to the noise in the PSD.
The PS2 of that filtered PSD is then computed and the region where the peaks associated with $\frac{1}{\dnu}$ and $\frac{2}{\dnu}$ are expected is once again selected. 
 Assuming that the modes frequencies follow an asymptotic pattern with typical amplitudes and widths, the properties of the Fourier transform lead to a higher peak at $\frac{2}{\dnu}$ than at $\frac{1}{\dnu}$ in the PS2.
Thus, a Gaussian function is fitted to the highest peak providing a measure for $\frac{2}{\dnu}$. The standard deviation of the fitted Gaussian function is taken as the uncertainty over the $\frac{2}{\dnu}$ measure and that uncertainty is propagated on $\dnu$ following the usual formula: $u(\dnu)=\frac{\dnu^2}{2}u\left(\frac{2}{\dnu}\right)$, where $u(\dnu)$ and $u\left(\frac{2}{\dnu}\right)$ are the uncertainties on $\dnu$ and $\frac{2}{\dnu}$ respectively.

\subsection{Undersampled super-Nyquist stars}
When a star is undersampled, i.e. some mode frequencies are super-Nyquist, its oscillation spectrum is folded around the Nyquist frequency. This yields a mirrored spectrum that in some cases overlaps with sub-Nyquist non mirrored modes. As stated in Section~\ref{sec:sn detection}, assuming that the true $\numax$ is sufficiently far beyond the Nyquist frequency, the modes observed in the mirrored PSD will not overlap with the modes existing below $\nuny$ because their height will be negligible compared to the mirrored ones (as in Fig. \ref{fig:supnyvsny} (b) ). We have realized that when the star is undersampled and $\numax-\frac{\text{FWHM}}{2}>\nuny$, the overlapping effect of the modes is negligible (where FWHM stands for Full Width at Half Maximum and is the width of the Gaussian envelope of the modes at half its maximum). Using the scaling relation $\text{FWHM}=0.25\numax$ \citep{2017ApJ...835..172L}, this condition is equivalent to having $\numax>323$~µHz. We then classified the stars with $\numax>320$~µHz as super-Nyquist, those with $\numax$ between $250$~µHz and $320$~µHz as close-to-Nyquist stars and those under $250$~µHz as sub-Nyquist stars.

As previously explained in  Section~\ref{sec:sn detection}, 
we test which pair of the set \{($\dnu$,$\nu_{\mathrm{max,measured}}$),($\dnu$, $2\nuny-\nu_{\mathrm{max,measured}}$)\} follows the best the scaling relation $\dnu\propto\numax^{0.77}$. That is used to identify the true $\numax$, whether it is the one mirrored around $\nuny$ or the directly measured one $\nusub$. This determines the final classification as sub-Nyquist, close-to-Nyquist, or super-Nyquist. The whole algorithm is visually described by a flowchart in appendix \ref{app:flowchart}.

To quantify the apodization effects (cf. appendix \ref{app:apo}) on the super-Nyquist measurement of $\numax$, we extract the seismic properties of three real \emph{Kepler} subgiants observed in SC that we use to simulate LC observations. Then, we follow the procedure described below:
\begin{itemize}
    
    \item For each of the three stars we fit their modes in the PSD of the SC dataset using the Python library \texttt{apollinaire} \citep{2022apo1,2023apo2} and the python tool \texttt{iechelle} \footnote{\url{https://github.com/dinilbose/iechelle/tree/main}}. Since the SC timeseries have a Nyquist frequency of $8\,496$ µHz, the apodization effect in the PSD of a star with $\numax$ between $300$ and $400$ µHz is negligible (indeed the change in the apodization function on that frequency range is very small $\sinc^2\left(\frac{300\pi}{2\times 8\,496}\right)-\sinc^2\left(\frac{400\pi}{2\times 8\,496}\right)\approx 0.001\ll1$).
    \item With the fitted mode parameters, we simulate a lightcurve with $20\times 1\,440$ days\footnote{$1\,440$ days being the duration of the \textit{Kepler} mission} using a sampling period of $1$ min and the same noise level as the original \emph{Kepler} observations. For each star we perform Monte Carlo simulations with 500 different noise realisations.
    \item From that SC lightcurve, we extract a 4-year long rebinned one at the \emph{Kepler} long-cadence rate of 29.4 minutes.
    \item Finally, we run PyA2Z on both the SC and the rebinned LC lightcurves and we compare the results.
\end{itemize}

The MC simulation showed a systematic effect of underestimating $\numax$ in the super-Nyquist regime, which is consistent with what we expect from theory. As the apodization function $\eta^2$ is a decreasing function on the range $[0,566]$ µHz, we expect the Gaussian envelope of the modes to be skewed towards the lower frequencies below its mean value. This result also confirms that the error bar of $\frac{\dnu}{2}$ is appropriate because, for more than  $99.93$\% of the realizations, the extracted $\numax$ of both cadences agree within their uncertainties.

\subsection{Handling close-to-Nyquist stars}
\label{sect:closeNyq}

For close-to-Nyquist stars, the measurement of $\numax$ is more complicated. The envelope of the modes consists of 2 overlapping Gaussian envelopes, one corresponding to the sub-Nyquist modes and the other to the mirrored super-Nyquist modes.

In this context, it is crucial to distinguish the real modes from the mirrored ones to mask them and measure $\numax$. To do so, the method we use is based on a straightening of the $\ell=0$ ridge of the echelle diagram. Given that the operation of symmetry does not affect the periodicity in $\dnu$ of the spectrum, a first measure of $\dnu$ is inferred from the PSD by the method described in Sect. \ref{dnu}.

\begin{figure}[h!]
     \centering
     \begin{subfigure}[b]{\linewidth}
         \centering
         \includegraphics[width=\linewidth]{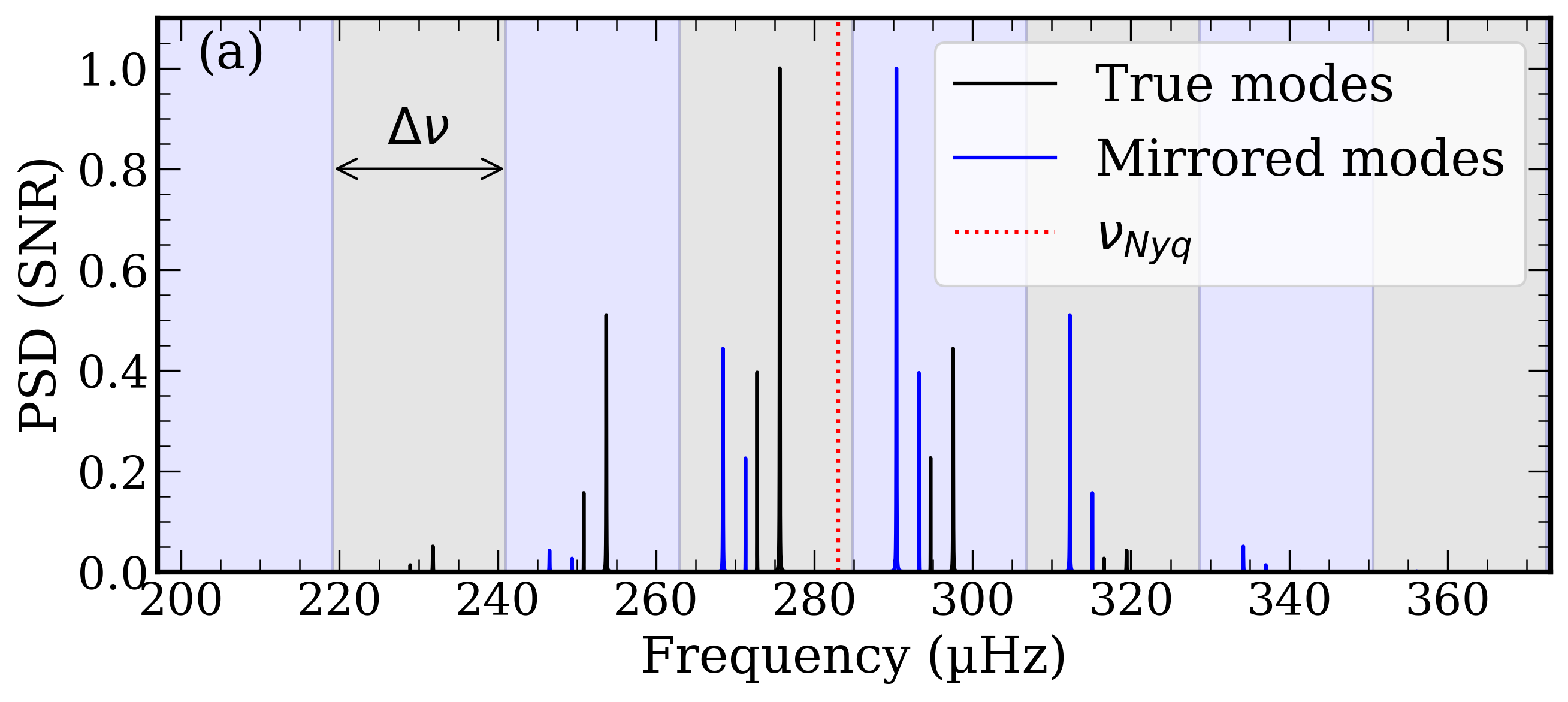}
         \label{fig:Synth_Nyq}
     \end{subfigure}
     \hfill
     \begin{subfigure}[b]{\linewidth}
         \centering
         \includegraphics[width=\linewidth]{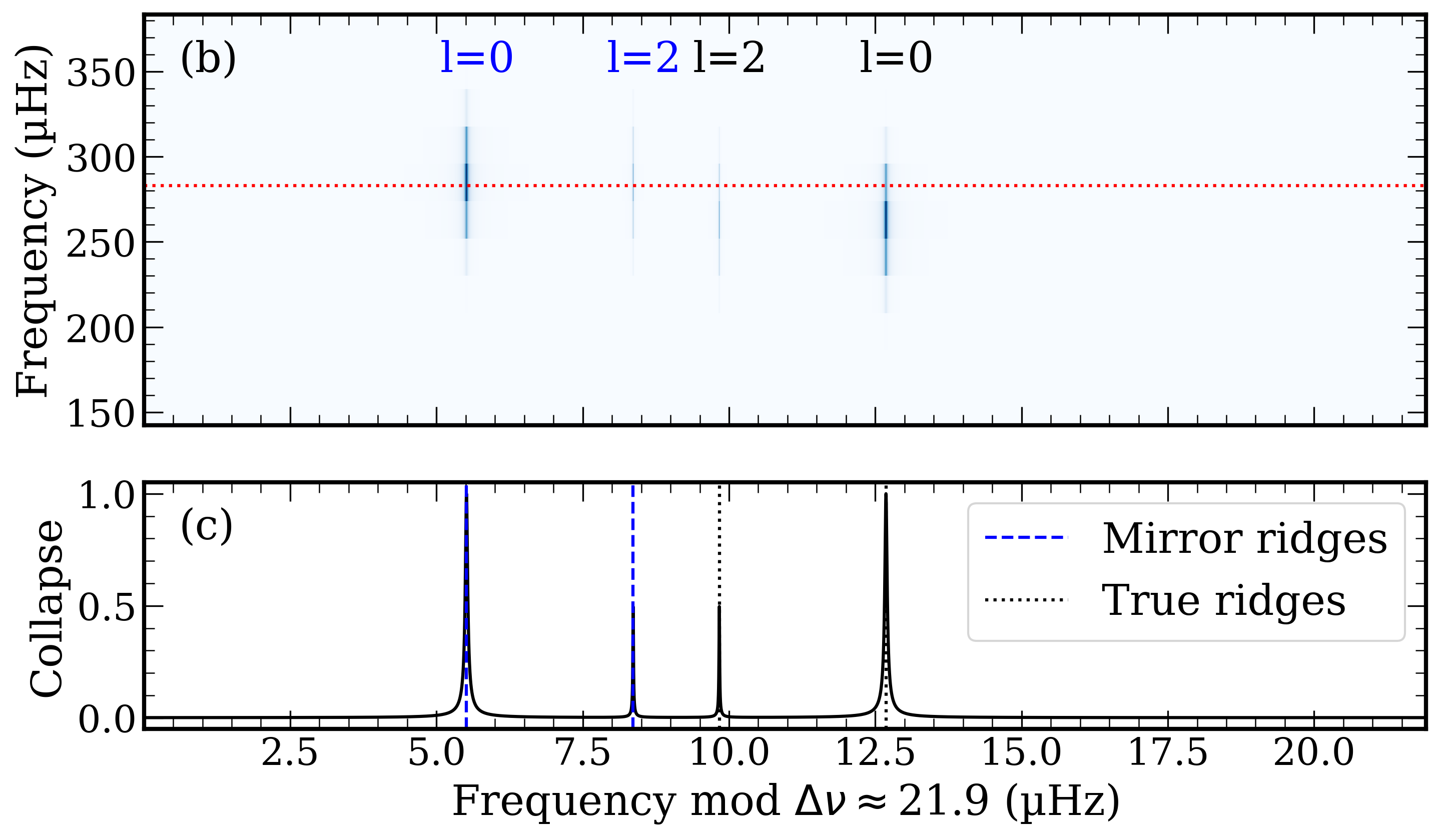}
         \label{fig:Nyq_ech}
     \end{subfigure}
        \caption{(a) Synthesized PSD of the $\ell=0,2$ modes of a star with $\numax=275\text{ µHz}$. The black modes are the true modes whereas the blue modes are those observed because of the undersampling: they are the black modes mirrored around $\nuny$.
        Each box of width $\dnu$ matches to one line of the echelle diagram shown in panel (b) along with its collapse in (c). In (b), the ridges due to the true modes are annotated in black whereas those annotated in blue are due to the shadow modes. In (c) the peaks due to the mirrored modes are marked with a blue vertical dashed line and the peaks due to the true modes are marked with a black vertical dotted line.}
        \label{fig:Nyq_exp}
\end{figure}

To construct the echelle diagram, the PSD is mirrored above $\nuny$ and concatenated with itself as shown in Fig. \ref{fig:Nyq_exp} (a), using only $\ell=0$ and 2 modes. This ensures that, once the echelle diagram is constructed, both ridges due to the real modes and those due to the mirrored modes are present and aligned as shown in Fig. \ref{fig:Nyq_exp} (b).

The straightening algorithm consists of varying the assumed $\dnu$ around the initial guess to maximize the amplitude of the $\ell=0$ peaks (due to the shadow and true ridges) in the collapse of the echelle diagram. Once that maximization is done, the algorithm proceeds to a peak detection and finds the candidates for the $\ell=0$ and $\ell=2$ ridges. The correct pair of ridges is found by searching for the set of peaks in the collapsed echelle diagram following the scaling relation $\delta\nu_{0,2}\approx 0.123\dnu$ in \cite{2012ApJ...757..190C_d02}.

Once that detection is done, the true $\ell=0$ modes are isolated in the PSD, the apodization is corrected by dividing the amplitudes of the modes by $\eta^2$ (cf. appendix \ref{app:apo}) and $\numax$ is measured again. The uncertainty on $\numax$ is once again given by $\frac{\dnu}{2}$. However, since the method straightens the ridges of the echelle diagram to get a measure of $\dnu$, the optimization process makes it challenging to compute a reliable value of the uncertainty on $\dnu$. To be conservative, we set a 10\% relative uncertainty on $\dnu$, corresponding approximately to the maximum uncertainty found in this work for the sub- and super-Nyquist samples.

\subsection{Global seismic parameters}
\label{sect:results}

\begin{table*}[]
\centering
\caption{Number of stars analyzed in each category.}
\begin{tabular}{cc c|c}
\hline\hline
Sample of origin& Main sample & Spectroscopic sample & Total \\ \hline
Sub-Nyquist                                   & 862                & 6                    & 868   \\ 
Super-Nyquist                                 & 284                & 1                    & 285   \\ 
Close-to-Nyquist (with $\dnu$ and $\numax$)   & 168               & 0                    & 168   \\ 
Close-to-Nyquist (without $\dnu$ nor $\numax$) & 423                & 0                    & 423   \\ 
Dubious                                       & 32               & 4                   & 36    \\ 
No $\dnu$                                     & 20                & 0                    & 20    \\ 
No signal                                     & 128               & 81                  & 209   \\ 
Spikes                                        & 17              & 11                   & 28    \\ 
Contamination                                 & 14                & 3                    & 17    \\ 
Pipeline failure                              & 11               & 0                    & 11    \\\hline 
Total                                         & 1959               & 106                  & 2065  \\ \hline
\end{tabular}

\label{tab:res}
\end{table*}

  Among the 2\,065 stars of our sample defined in Sect.~\ref{sec:Data preselection}, our pipeline PyA2Z presented in Sect.~\ref{sec:sn detection} along with a visual inspection, allowed us to reclassify 285 stars as super-Nyquist (see Table \ref{tab:res} and Table \ref{supnytable}) and measured the global seismic parameters of 168 close-to-Nyquist stars (cf. Table \ref{tab:res} and Table \ref{nytable}).

\begin{table}[h!]
\centering
\caption{Global seismic parameters from the confirmed 285 super-Nyquist stars.}
\begin{tabular}{ccccc}
\hline\hline
KIC & $\numax$ & $\dnu$ & $\text{Err}_{\dnu}$ & $\text{Err}_{\numax}$ \\
&(µHz)&(µHz)&(µHz)&(µHz)\\
\hline
1164274 & 398.98 & 25.75 & 2.56 & 12.87 \\
1296817 & 335.52 & 22.92 & 1.13 & 11.46 \\
1866174 & 362.72 & 23.66 & 1.19 & 11.83 \\
2141652 & 333.19 & 21.97 & 1.52 & 10.99 \\
2142186 & 327.83 & 22.25 & 1.14 & 11.13 \\
\multicolumn{5}{c}{$\cdots$}\\
8442463 & 331.95 & 21.10 & 0.85 & 10.55 \\
8543816 & 362.25 & 25.43 & 2.04 & 12.72 \\
8545556 & 334.42 & 23.42 & 0.98 & 11.71 \\
8672397 & 326.89 & 23.01 & 1.36 & 11.5 \\
8674803 & 338.73 & 22.76 & 1.25 & 11.38 \\
\hline\\
\end{tabular}

\label{supnytable}
\tablefoot{The full table is available at the CDS.}
\end{table}

\begin{table}[h]
\centering
\caption{Global seismic parameters from the confirmed 168 close-to-Nyquist stars.}
\begin{tabular}{cccc}
\hline\hline
KIC & $
\nu_{\mathrm{max}}$ & $\Delta
\nu$ & $\mathrm{Err}_{
\nu_{\mathrm{max}}}$ \\
&(µHz)&(µHz)&(µHz)\\
\hline
10910840 & 276.23 & 18.85 & 9.42 \\
8586548 & 261.68 & 18.86 & 9.43 \\
11605794 & 286.55 & 19.12 & 9.56 \\
11709205 & 274.70 & 19.13 & 9.57 \\
10990544 & 278.27 & 19.09 & 9.55 \\
\multicolumn{4}{c}{$\cdots$} \\
11138444 & 294.36 & 20.46 & 10.23 \\
7340867 & 302.05 & 21.15 & 10.57 \\
8570418 & 260.46 & 18.83 & 9.42 \\
9396463 & 275.14 & 19.68 & 9.84 \\
12057675 & 264.48 & 18.79 & 9.39 \\
\hline
\end{tabular}

\label{nytable}
\tablefoot{The full table is available at the CDS. No uncertainties are reported on $\dnu$ owing to the optimization method used in close-to-Nyquist stars.}
\end{table}

Among the super- and sub-Nyquist stars, 309 lack global seismic parameters (taking into account the main and the spectroscopic samples). These can be explained by different factors such as the absence of a detectable signal above 150 µHz in the PSD (labeled as 'No signal' in Table \ref{tab:res}), a failure of the pipeline to determine $\Delta\nu$ (labeled as 'No $\dnu$' in Table \ref{tab:res}), spikes (usually stemming from a companion object or instrumental effects) that could not be properly removed (labeled as 'Spikes' in Table \ref{tab:res}), contamination in the lightcurve (labeled as 'Contamination' in Table \ref{tab:res}), and the pipeline’s inability to converge during the fitting process (labeled as 'Pipeline failure' in Table \ref{tab:res}).
36 stars of the sample have a detection of $\numax$ and $\dnu$ for which we are unsure of the correctness of the result (because the visual inspection of the echelle diagrams could not confirm the detection), they are labeled as 'Dubious' in Table~\ref{tab:res}). Hence, they are excluded from the final table.

Among the 591 close-to-Nyquist stars, our pipeline successfully provided measurements of $\dnu$ and $\numax$ for 28\% of them. 

 Most of the initial pipeline's misdetections were due to a measurement of the second harmonic of the $\dnu$ periodicity in the PS2 (peak corresponding to $\frac{3}{\dnu}$) instead of the first (peak at $\frac{2}{\dnu}$). This problem, which was observed several times upon visual examination of the measurements, was most prominent when the power in the $\ell=1$ was not as high as expected \citep[for instance in the PSDs of depressed dipole mode stars, e.g.][]{2014A&A...563A..84G}. This phenomenon arises when the amplitudes of the $\ell=1$ modes are reduced, the dominant periodicity of the PSD asymptotically becomes $\dnu$ and we no longer observe the significantly greater peaks at $\frac{2k}{\dnu},\, k \in\mathbb{N}$ harmonics as expected in the case of a $\frac{\dnu}{2}$ dominant periodicity of the PSD.
 
 This observation led to the implementation of a test for flagging depressed dipole modes stars in our dataset: if the ratio of the heights of the peaks of 2 consecutive peaks at $\frac{2k+1}{\dnu}$ and $\frac{2k}{\dnu}$, $k\in\mathbb{N}$ of the PS2 is close to 1, the star is flagged and a visual inspection is made to confirm whether the dipole modes are indeed depressed or not.

Among the 47 stars with $\numax$ above the Nyquist frequency in \citet{2016MNRAS.463.1297Y} we confirmed the global seismic parameters of 46 of them. 

However, for KIC~4759008, our analysis --including a visual inspection-- cannot confirm the published value.

 Our methodology increased the sample of stars with $\numax>283$ $\mu$Hz (which includes super-Nyquists stars as well as some close-to-Nyquist stars) by a factor of 7. The stars we have detected along with the scaling relation we used are plotted in appendix \ref{app:scalrel}.

 Since TESS observations have shorter cadences compared to {\it Kepler} after cycle 3\footnote{\url{https://heasarc.gsfc.nasa.gov/docs/tess/data-products.html}} (10 and 3 minutes respectively for cycles 3 and 4 and from cycle 5 onwards), our \emph{Kepler} stars are not super-Nyquist in these TESS data and could be analyzed using traditional asteroseismic methods. Thus, we have studied all our stars that have been observed in the two TESS extended missions and for which the Quick Look Pipeline \citep[QLP,][]{2020RNAAS...4..204H,2020RNAAS...4..206H,2022RNAAS...6..235K} data are available (see appendix~\ref{app_tess}). Unfortunately, most of the \emph{Kepler} stars are too faint to have measurable oscillations in TESS. We were able to perform the seismic analysis in only 9 stars (see Fig. \ref{fig:TESSVSKEP}). For all of them, we measured a $\numax$ from TESS that is consistent with our claimed $\numax$ value from \textit{Kepler} (see Table~\ref{Tab:Tess_results}). Given that this sample includes two stars that we identified as Super-Nyquist, this gives us additional confidence that our methods are correctly obtaining the true $\numax$ from the \kep data.

\section{Fundamental stellar parameters}
\label{sec:MRA}

Asteroseismology, Gaia data, and spectroscopic surveys are a powerful combination. Indeed, precise and accurate Gaia luminosities, combined with $\teff$, provides an independent measure of radius that can be used to test asteroseismic scaling relations. The asteroseismic scaling relations require knowledge of $\teff$ directly. Mapping $\dnu$ to mean density and age requires knowledge of the heavy element abundance mixture. We therefore begin by summarizing the spectroscopic data that we have used, and follow with our method for inferring mass, radius, and age.

\subsection{Atmospheric parameters}
\label{sec:spectra}

High resolution spectroscopy is uniquely capable of measuring stellar abundances. We adopt [Fe/H] and $[\alpha/\mathrm{Fe}]$ to characterize the abundance pattern relative to a reference solar mixture (see APOKASC-3 for a discussion). Our stellar models and inferences use the \citet{1998SSRv...85..161G} heavy element mixture for the Sun, which is in good agreement with helioseismic data \citep{basinger24}. Spectroscopic $\teff$ data can be tied to an absolute scale, and these $\teff$ estimates are less susceptible to uncertainties arising from interstellar extinction than $\teff$ inferred from characterizing the spectral energy distribution. Spectroscopic data are therefore highly desirable for precise and accurate stellar characterization \citep{2010A&A...512A..54C,gonzalezhernandez09}. 

APOGEE is part of the Sloan Digital Sky Survey (SDSS; \citealt{York2000}), which has now gone through 5 major surveys and 17 data releases. Data in the \kep fields was obtained during SDSS-III \citep{sdss3} and SDSS-IV \citep{sdss4}. APOGEE is a multi-fiber survey that uses a high resolution ($R\sim 22,000$) infrared spectrograph \citep{Wilson2019} mounted on the SDSS telescope \citep{Gunn2006}. APOGEE targeting, described in \citet{zasowski_2013,zasowski_2017, Beaton2021}, focused on the \kep fields largely because of the overlap with asteroseismic targets. Stellar parameters are inferred using a multi-dimensional chi-squared analysis technique, the ASPCAP pipeline \citep{aspcap}, and are tied to a fundamental scale in a post-processing step. The $\teff$ scale is the \citet{gonzalezhernandez09} calibration of the infrared flux method. We use the latest public data, Data Release 17 \citep[hereafter DR17,][]{DR17}. Median uncertainties in $\teff$, [Fe/H], and $[\alpha/\text{Fe}]$, as discussed in \citet{apokasc3}, are 42K, 0.05 dex, and 0.05 dex respectively. These data are also our sole source of [C/Fe] and [N/Fe], important for testing the first dredge-up (FDU) of CNO cycle processed material in red giants (see below).

We do not have spectra for all of our targets, however, and as a result we need to expand our methodology to include information from other sources. Atmospheric parameters from APOGEE DR17 are used in priority when available, and we complete the sample with the {\it Gaia} XP spectra ($R\sim100$; \citealt{carrasco21}). These data have been used to infer abundances and $\teff$ for 175 million stars, using APOGEE DR17 as the training set for the XGboost algorithm \citep{chen16}, leading to the catalog by \citet{2023ApJS..267....8A}. For the 1321 stars that we classified as sub-, close-to-, and super-Nyquist (see section~\ref{sect:results}), 640 stars have APOGEE parameters and 1249 stars have XGBoost values with an overlap for 611 stars. With this, we are left with 43 stars that lack spectroscopic and spectro-photometric parameters.

\subsection{Computing the stellar fundamental parameters}
\label{sec:Aldo}

The computation of stellar masses, radii and ages is done with \texttt{BeSPP} \citep{Serenelli:2013,2017ApJS..233...23S} and it follows the methodology used for the APOKASC-3 catalog of evolved stars \citep{apokasc3}. We use a grid of stellar models computed with \texttt{GARSTEC} \citep{weiss:2008} that  spans the mass range between 0.6 and 5$\msun$, and $-2.5$ and +0.6 for [Fe/H]. It is the same grid used in \citet{apokasc3} and a detailed description of the physical inputs of the stellar models can be found in that work. Spectroscopic stellar parameters are adopted from APOKASC-3 where available, and otherwise from XGBoost (see sec.~\ref{sec:spectra}).

Seismic quantities for each model of the grid are computed as follows. $\numax$ is computed from the scaling relation 
\begin{equation}
\frac{\numax}{\nu_{\mathrm{max}, \odot}} = \left(\frac{g}{{\rm g}_\odot}\right)\left(\frac{\teff}{\teffsun}\right)^{-1/2},
\label{eq:numax}
\end{equation}
where g and $T_\mathrm{eff}$ are the surface gravity and effective temperature of the models.
$\dnu$ is computed as the slope of a linear fit to the frequencies of radial modes with frequencies lower than the acoustic cutoff frequency. For the fit, frequencies are weighted assuming a Gaussian envelope centered in $\numax$ with a FHWM given by $0.66\numax^{0.88}$ \citep{mosser:2012}. The ratio
\begin{equation}
\fdnu = \frac{\Delta \nu_{\odot}}{\dnu} \left(\frac{M}{\msun}\right)^{1/2} \left(\frac{R}{\rsun}\right)^{-3/2}
\end{equation}
measures the deviation of stellar models away from a strict scaling relation between $\dnu$ and the square root of the mean stellar density. From the relations above, mass and radius can be formally obtained from
\begin{equation}
\frac{M}{\msun} = \left( \frac{f_{\nu_\mathrm{max}}\nu_{\rm max}}{\nu_{\rm max, \odot}} \right)^3 \left( \frac{f_{\Delta \nu} \Delta \nu}{\Delta \nu_{\odot}} \right)^{-4}
\left( \frac{T_{\rm eff}}{\teffsun} \right)^{1.5}
\label{eq:mass}
\end{equation}
and
\begin{equation}
\frac{R}{\rsun} = \left( \frac{f_{\nu_\mathrm{max}}\nu_{\rm max}}{\nu_{\rm max, \odot}} \right) \left( \frac{f_{\Delta \nu} \Delta \nu}{\Delta \nu_{\odot}} \right)^{-2}
\left( \frac{T_{\rm eff}}{\teffsun} \right)^{0.5}.
\label{eq:radius}
\end{equation}
We have chosen solar reference values to be consistent with APOKASC-3 (${\nu_{\mathrm{max, BeSPP}, \odot}}$=3076 µHz and $\Delta \nu_{\odot}$=135.146 µHz) and we have used a constant $f_{\nu_\mathrm{max}}=0.996$: the correction to the solar $\numax$ computed in APOKASC-3 for stars with $\numax>50\text{ µHz}$ \citep{apokasc3}. 

The first step in the calculation of stellar fundamental parameters is to use $\dnu$, $\numax$ and $\teff$ in Eqs.~\ref{eq:numax} and \ref{eq:mass} to compute $g$ and $M$, assuming $\fdnu=1$, i.e. perfect scaling relations. We use $M$, $g$, and [Fe/H]$_{\rm corr}$\footnote{[Fe/H]$_{\rm corr}$ = [Fe/H] + 0.625 $\alphafe$ is used to account for alpha enhancement, as the grid of stellar models uses a solar-scaled composition. This is the same correction as the one adopted in \citet{apokasc3} and is quantitatively similar, to other corrections, such as the one from \citet{salaris_chieffi_straniero1993}.} as input quantities in \texttt{BeSPP} to produce a posterior distribution of $\fdnu$ and then an updated stellar mass using Eq.~\ref{eq:mass}. The process is continued iteratively until convergence for $\fdnu$ is obtained when changes in $\fdnu$ are smaller than one part in 10$^5$, typically requiring 3 to 5 iterations. The final seismic stellar mass and radius, and their uncertainties, are then obtained using Eq.~\ref{eq:mass} and \ref{eq:radius} one last time. We note that, since the XGBoost dataset does not include measurements of [$\alpha$/Fe], we adopted [$\alpha$/Fe] = 0 when using this data. To account for the additional uncertainty introduced by this assumption, we inflated the reported uncertainties from 0.1 dex to 0.15 dex in metallicity and from 50 K to 70 K in effective temperature.

To determine the age, we run \texttt{BeSPP} again, using $g$, [Fe/H]$_{\rm corr}$ and the seismic mass as inputs. However, the age dependence on stellar mass departs from a local linear behavior when mass uncertainties are not small. In order to provide age central values and uncertainties that are consistent with those of the seismic masses, we proceed as follows. \texttt{BeSPP} is run three times using as input in each of the runs, respectively: the central seismic mass, the central mass increased by its uncertainty, and the central mass decreased by its uncertainty. 
For each of the three runs, a small formal mass uncertainty of 0.01 M$_\odot$ is used; this does not reflect the true uncertainty, which is instead accounted for by considering the spread across the three runs.
The first run is used to determine the central value of the age of the star and the standard deviation is assumed to represent uncertainties linked to $g$ and [Fe/H]$_{\rm corr}$ uncertainties. The other two runs are used to compute the $-1\sigma$ and $+1\sigma$ age uncertainties, respectively, after adding quadratically the standard deviations from the central run.

\subsection{Comparison with Gaia}
In order to vet the asteroseismic results, we compare to alternate radius and mass scales using Gaia radii. We use Gaia DR3 parallaxes \citep{gaia_collab_23,lindegren+2021a} as the basis for a luminosity scale, corrected according to \cite{lindegren+2021b} and with non-single stars removed as well as spurious parallax solutions \citep{rybizki+2022}. These parallaxes, when combined with extinctions, 2MASS K-band photometry and bolometric corrections, as well as spectroscopic temperatures, will yield a radius; this approach has been detailed in \cite{zinn+2017}. As in Sec.~\ref{sec:Aldo} spectroscopic stellar parameters are adopted from APOKASC-3 where available, and otherwise from XGBoost. The K-band bolometric correction is interpolated from MIST \citep{choi+2016a,dotter+2016a,paxton+2011,paxton+2013a,paxton+2015a} bolometric correction tables in metallicity, gravity, and temperature. The tables are constructed using the C3K grid of 1D atmosphere models (C. Conroy et al., in preparation; based on ATLAS12/SYNTHE; \citealt{kurucz1970,kurucz1993}). Salaris-corrected metallicities are adopted \citep{salaris_chieffi_straniero1993} when $\alphafe$ is available from APOKASC-3 and otherwise we assume $\alphafe = 0$ with an uncertainty of $0.3$dex. Finally, K-band magnitudes are required to have an `A' quality rating from 2MASS. We use asteroseismic surface gravities for the bolometric calculation, and therefore we only compare to stars with measured $\numax$. Extinctions are calculated using a three-dimensional dust map based on \cite{marshall+2006}, \cite{green+2019} and \cite{drimmel_cabrera-lavers_lopez-corredoira2003}, as implemented in \texttt{mwdust} \citep{bovy+2016}. $\dnu$ values are corrected with the $\fdnu$ scale from \texttt{asfgrid} \citep{sharma+2016,stello_sharma2022}. For this purpose, evolutionary states are taken to be RGB unless categorized as `RC' in APOKASC-3.

\begin{figure}[h!]
    \centering
    \includegraphics[width=\linewidth]{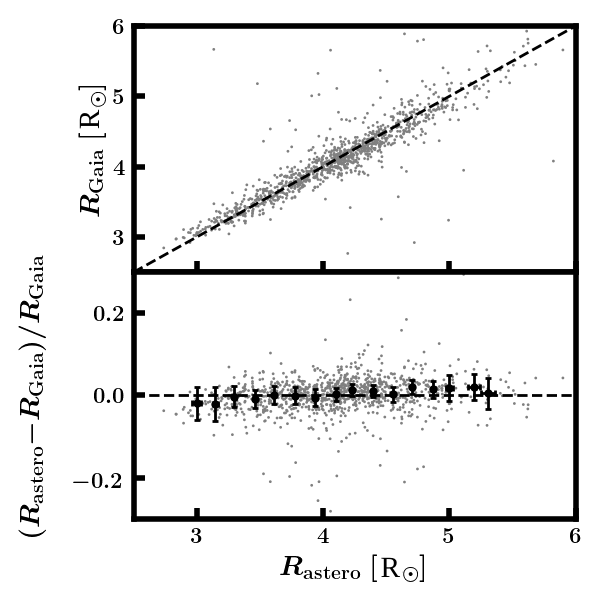}
    \caption{One-to-one comparison of Gaia and asteroseimsic radius scales (top) and the fractional differences (bottom; with binned medians and uncertainties on the median). The two show excellent agreement (to within $0.9\%$ on average), which demonstrates the accuracy of the asteroseismic analysis here and of the asteroseismic radius scale in the subgiant regime.}
    \label{fig:radius_comp}
\end{figure}

Fig.~\ref{fig:radius_comp} shows excellent agreement between asteroseismic and Gaia radii --- to within $0.9\%$ on average. The sample presented here lies between the two radius regimes in which \citep{Zinn2019Rtest} quantified the asteroseismic and Gaia radius scale agreement. In that work, stars with $R < 3.5\rsun$ were found to have a ratio $R_{\mathrm{astero}}/R_{\mathrm{Gaia}} = 0.979 \pm 0.005$ and stars with $10 \rsun < R < 30\rsun$ were found to have a ratio $R_{\mathrm{astero}}/R_{\mathrm{Gaia}} = 1.019\pm 0.006$. In our intermediate radius regime, the agreement lies between these values, suggesting a smooth transition in systematics --- either related to scaling relations or perhaps temperature scales --- between the dwarfs and giants. (See also Fig. 3 of \citealt{Zinn2019Rtest}.)

This radius comparison constrains a degenerate combination of $\fnumax$ and $\fdnu$ since the asteroseismic radius is proportional to $\fnumax/\fdnu^2$. However, we can also test the $\fnumax$ and $\fdnu$ scale separately by constructing single--scaling relation masses (which will be henceforth referred to as Gaia masses):

\begin{equation}\label{eq:mdnu_scaling}
    \frac{M_{\dnu}}{\msun} = \bigg(\frac{f_{\Delta \nu}\Delta \nu}{\Delta \nu_{\odot}}\bigg)^2 \bigg(\frac{R_\mathrm{Gaia}}{\rsun}\bigg)^{3} 
\end{equation}

and

\begin{equation}\label{eq:mnumax_scaling}
    \frac{M_{\nu_\mathrm{max}}}{\msun} = \bigg(\frac{f_{\nu_\mathrm{max}}\nu_{\mathrm{max}}}{\nu_{\mathrm{max}, \odot}}\bigg) \bigg(\frac{R_\mathrm{Gaia}}{\rsun}\bigg)^{2} \bigg(\frac{\text{T}_{\text{eff}}}{\teffsun}\bigg)^{1/2}.
\end{equation}

Here, $f_{\nu_\mathrm{max}}$ and $f_{\Delta \nu}$ can both in theory be a function of stellar parameters, and, if not unity, would reflect systematic errors in the either measurement or the mapping of observed quantities to the theoretically-motivated scaling relations. Note that we adopt $\nu_{\mathrm{max,asfgrid}, \odot} = $3043 µHz appropriate for using \texttt{asfgrid} $f_{\Delta \nu}$ values \citep{apokasc3}.

\begin{figure}[h!]
    \centering
    \includegraphics[width=\linewidth]{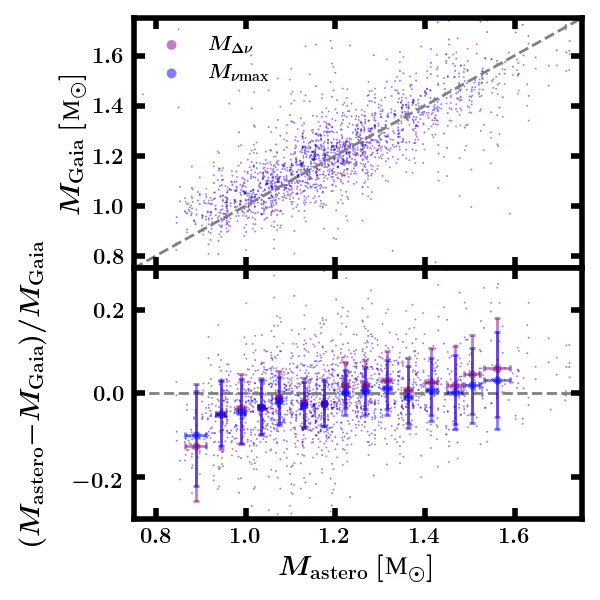}
    \caption{One-to-one comparison of Gaia and asteroseismic mass scales (top) and the fractional differences (bottom; with binned medians and uncertainties on the median). The mean agreement between the different mass scales, the Gaia mass scales $M_{\numax}$ and $M_{\dnu}$, and the asteroseismic mass scale, $M_{\mathrm{astero}}$, are consistent to within the uncertainties on the mean ($1.5\%$, respectively).}
    \label{fig:mass_comp}
\end{figure}

In Fig.~\ref{fig:mass_comp}, we look for evidence of such systematics in $\numax$ or $\dnu$ in the differences of the mass distributions that result from Eqs.~\ref{eq:mdnu_scaling}~\&~\ref{eq:mnumax_scaling}. The agreement is excellent among all three mass scales --- at the 1.5\% level. The agreement of the two Gaia mass scales with each other is even better than this, which can be seen in the nearly complete overlap between binned medians in Figure~\ref{fig:mass_comp}. 
This behavior can be understood analytically by examining the ratio of equations (\ref{eq:mdnu_scaling}) and (\ref{eq:mnumax_scaling}), which depends linearly on $R_\mathrm{Gaia}$. In contrast, the individual expressions (\ref{eq:mdnu_scaling}) and (\ref{eq:mnumax_scaling}) scale with $R_\mathrm{Gaia}^2$ and $R_\mathrm{Gaia}^3$, respectively. This indicates that they are more sensitive to potential systematics in the Gaia radius measurements.
\\
At this stage, we found that the scatter in stellar masses and radii—when compared to Gaia data—was smaller than the uncertainties obtained from propagating errors in $\dnu$ and $\numax$. To address this discrepancy, we decided to rescale our error bars. We divided the sample into bins of mass (or radius), within which we computed the scatter of the fractional differences (see Fig. \ref{fig:radius_comp}, \ref{fig:mass_comp}). We then rescaled the uncertainties in the masses (or radii) for each bin so that the mean relative uncertainty matched the observed scatter in the fractional differences.
On average this has decreased the uncertainty on the radius by a factor of 4 and decreased the uncertainty on the mass by a factor of 2.

\begin{figure*}[h!]
    \centering
    \includegraphics[scale=0.45]{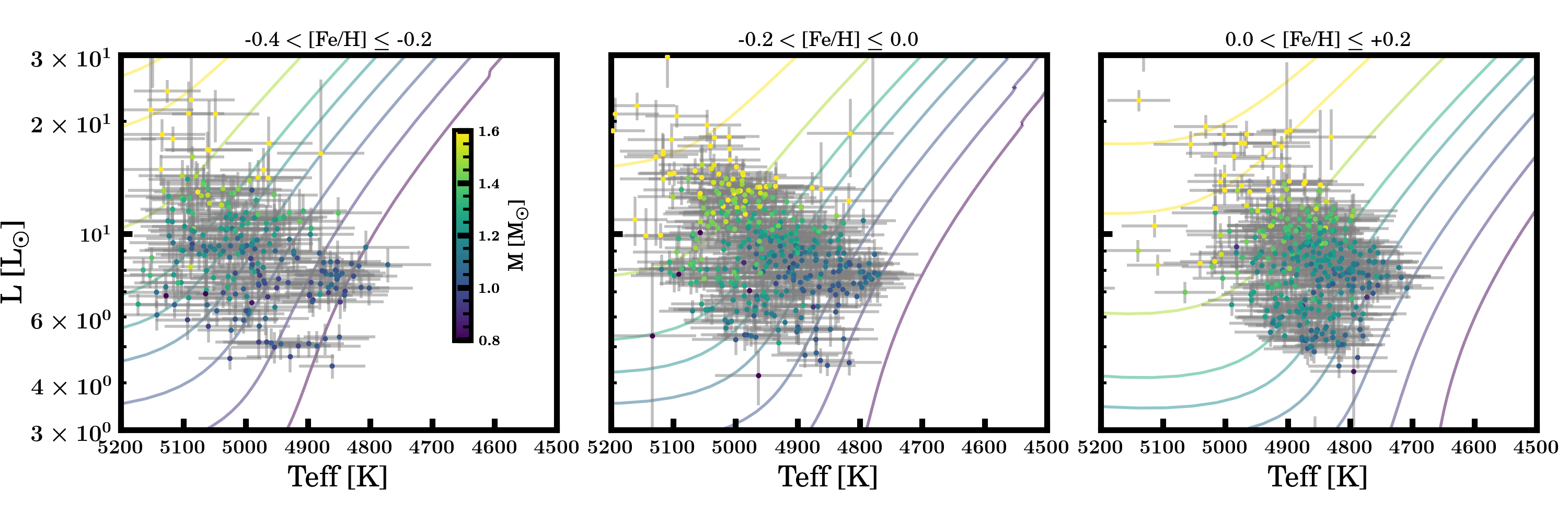} 
    \caption{HR diagram color-coded by $\numax$+Gaia mass and binned by metallicity. Luminosity is computed from Gaia radii. MIST evolutionary tracks for stars with masses between 0.8$\msun$ and 1.6 $\msun$ and metallicities equal to the central metallicity of each bin are shown for reference. The sample spans an interesting range in mass/age as well as metallicity for studying the first dredge-up (the clump is visible in the MIST tracks as kinks above the locus of the sample); see Fig.~\ref{fig:fdu}.}
    \label{fig:mass_hr}
\end{figure*}

\section{Discussion}
\label{sec:discussion}

\begin{figure*}[h!]
    \centering
    \includegraphics[width=\linewidth]{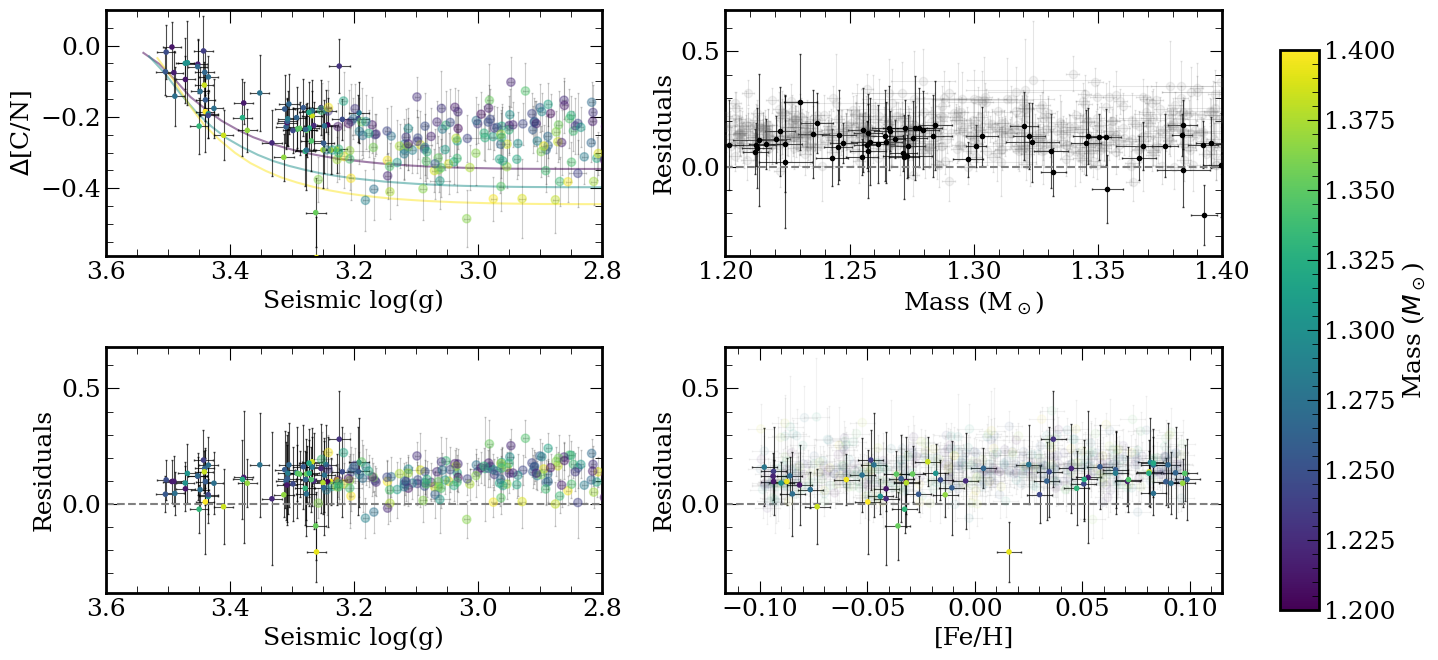}
    \caption{ $\Delta[\mathrm{C}/\mathrm{N}]$ \textit{i.e.} change in the surface carbon to nitrogen abundance ratio since birth VS the seismic $\log(g)$ and residuals as a function of $\log(g)$, mass and metallicity for stars of metallicity $-0.1<[\mathrm{Fe}/\mathrm{H}]<0.1$ and masses $1.2\mathrm{M}_\odot<M<1.4\mathrm{M}_\odot$ found in the present work (marked with opaque dots) along with stars found in the APOKASC-3 catalog (marked with semi-transparent circles) (\cite{apokasc3}). The stars flagged as being part of the red clump in APOKASC-3 were removed from this figure. Evolutionary tracks from \cite{2025arXiv250114867C} with solar metallicity are overlayed on top of the data. The errorbars on mass and metallicity have been shrunk of factors 20 and 10 respectively in order to improve readibility of the graph. Stars evolve from left to right on the left panels of the plot. We note that the end of the dredge-up is expected to be between $\log g=2.9$ and $\log g=3.1$ \citep[see Fig. 4 of ][]{2025arXiv250114867C} hence all our stars are undergoing the first dredge-up.}
    \label{fig:fdu}
\end{figure*}

Stars near the \kep Nyquist frequency are particularly interesting for constraining stellar models. As we show in the HR diagrams  Fig.~\ref{fig:mass_hr}, these stars sit in the curve where stars turn from moving horizontally across the subgiant branch to moving vertically up the red giant branch. Our sample spans a range in metallicity (-0.4 to +0.2 dex) and in mass (0.8 $\msun$ to 1.6 $\msun$). When we compare the measured temperatures and luminosities of these stars to the predictions of stellar models, we find that the stars in general are well matched to the models as a function of mass and metallicity, consistent with previous results (e.g., \cite{grusnis_2024}, submitted). Such agreement is not generally found for more evolved giants \citep[e.g.][]{2017ApJ...840...17T}. This agreement suggests that masses and ages estimated from temperature and luminosity are likely to be reliable in this regime, providing an opportunity for large galactic archaeology studies based on Gaia data \citep[e.g.,][]{2022Natur.603..599X, 2024ApJ...976...87N}.

\begin{figure}
    \centering
    \includegraphics[width=\linewidth]{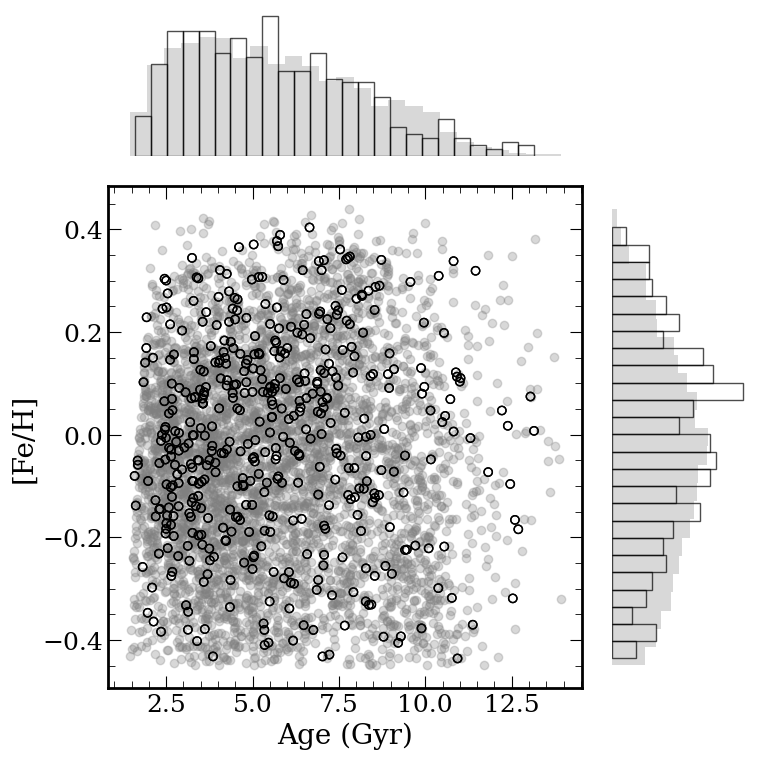}
    \caption{Metallicity vs. Age: The open black circles and black-outlined histograms represent the sample from this paper with available APOGEE spectra, while the semi-transparent circles and grey histograms correspond to data from the APOKASC-3 paper \citep{apokasc3}. Ages outside the range of 0 to 14 Gyr have been excluded. The stars flagged as being part of the red clump in APOKASC-3 were removed from this figure.}
    \label{fig:met_age}
\end{figure}

\begin{figure}
    \centering
    \includegraphics[width=\linewidth]{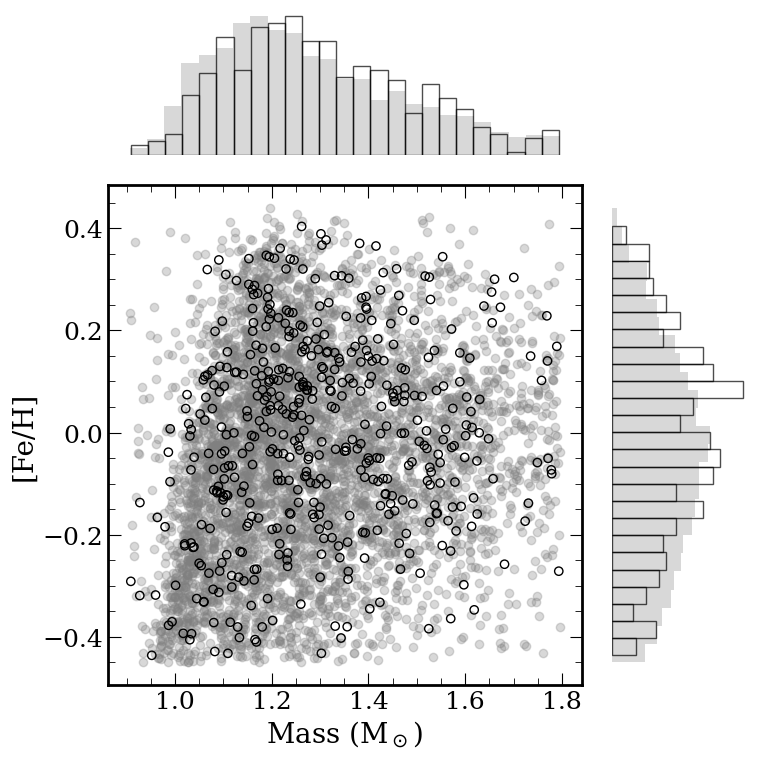}
    \caption{Mass vs. [Fe/H]: The open black circles and black-outlined histograms represent the sample from this paper with available APOGEE spectra, while the semi-transparent circles and grey histograms correspond to data from the APOKASC-3 paper \citep{apokasc3}.The stars flagged as being part of the red clump in APOKASC-3 were removed from this figure.}
    \label{fig:mass_met}
\end{figure}

The stars in this sample also cover the moment in which stars undergo the first dredge up. During this process, the convective envelope deepens into a region that has previously undergone the CNO cycle, and thus has a different balance of carbon and nitrogen than the stellar surface. This processed material is mixed up to the stellar surface, and changes the observed abundance ratios once it occurs. How much the ratios change depends on both the depth of the dredge up, and the amount of CNO burning that happened in that zone, and should therefore depend most on the mass and somewhat on the metallicity of the star. We show in Fig.~\ref{fig:fdu} the observed ratio of the carbon to nitrogen [C/N] as a function of surface gravity, compared to the predicted evolution in stellar models of various masses. In general, the location of the dredge up seems to be close to correct, but the depth of the dredge up, and its mass dependence in the data do not match the predictions of models very well. This could be due to errors in the assumed model physics, or to errors in the observed abundance scale, which is notoriously difficult to calibrate and varies between surveys \citep{2018AJ....156..126J}. This zero-point offset is also observed in \cite{2025arXiv250114867C},  e.g. in Fig. 6, where it is argued that these probably stem from the APOGEE data itself or from the assumption that trends in carbon and nitrogen abundance in subgiants \citep{2024MNRAS.530..149R} and the birth trends of RGB stars are the same. In either case, this sample could prove uniquely valuable for constraining the mixing occurring during the first dredge up, expanding the work done by \cite{2024MNRAS.530..149R}.
\\
This work successfully reproduces, at a qualitative level, the population trends observed in \cite{apokasc3}, as illustrated in Figs.~\ref{fig:met_age} and \ref{fig:mass_met}. In particular, Fig.~\ref{fig:mass_met} reveals the same population effect as Fig.~18 in \cite{apokasc3}, namely the absence of low-metallicity, low-mass stars in our sample.

While broadly similar, the distributions of mass, age, and metallicity in our sample and the APOKASC-3 catalog exhibit some differences:
\begin{itemize}
\item A noticeable under-representation of stars aged between 9 and 10.5 Gyr. This is likely due to an under-representation of alpha rich stars in our sample compared to the APOKASC-3 one. Indeed, plotting the same figure as Fig. \ref{fig:met_age} with only alpha poor stars gets rid of this effect. However, given that the age uncertainties are typically larger than 2 Gyr, the statistical significance of the dip observed in the histogram remains uncertain.
\item Fewer stars with low [Fe/H] compared to the APOKASC-3 sample.
\item A higher number of stars with [Fe/H] between 0.1 and 0.2 than in the APOKASC-3 catalog.
\item A lower number of low-mass stars compared to APOKASC-3.
\end{itemize}
These differences likely stem from the smaller size of our sample relative to the APOKASC-3 catalog.

\section{Conclusion}
\label{sect:conclusion}

In this paper, we developed part of the PyA2Z pipeline, a revised python version of the A2Z pipeline. In addition to the old pipeline, PyA2Z flags stars that are likely to be super-Nyquist. This methodology relies on the seismic scaling relations, where  the measured $\dnu$ is compared with the $\numax$ obtained below $\nuny$ and with $\numax$ from the mirrored PSD.  

We have demonstrated that PyA2Z is able to identify and correctly measure stars whose oscillations are close to or just above the Nyquist frequency. We lay out a method for distinguishing real from mirrored modes by identifying the radial-quadrupole structure in the echelle diagram that follows the seismic scaling relation $\delta\nu_{0,2}(\dnu)$ calibrated for red-giant stars. 

Applying these methods on a sample of 2\,065 \kep\, targets yields a new asteroseismic catalogue of 285 super-Nyquist stars as well as 168 close-to-Nyquist stars for which we give the global seismic parameters $\dnu$ and $\numax$. Our results expanded the known sample of super-Nyquist oscillators by a factor of 7.

Using a grid of stellar models computed with GARSTEC, we obtained the masses, radii, and ages of 892 stars. Our sample covers a range in mass of 0.8 to 1.6 $\msun$ and in metallicity from -0.4 to +0.2 dex. Comparisons of measured temperatures and luminosities of our set of stars with stellar models show good agreement in this evolutionary phase, suggesting that mass and age estimates from temperature and luminosity are reliable in this regime.

The study of the dredge-up phase shows that its timing matches model predictions, but the depth and mass dependence do not align well. This could be due to uncertainties in model physics or calibration issues in observed abundance scales. Our study qualitatively reproduces trends observed in previous works but reveals key differences when compared to the APOKASC-3 catalogue. Specifically, our sample shows an under-representation of stars aged between 9 and 10.5 Gyr, as well as a lower fraction of metal-poor stars. Conversely, we observe a higher proportion of stars with slightly enhanced metallicity ($[\mathrm{Fe}/\mathrm{H}]$ = 0.1–0.2) and a lower number of low-mass stars relative to APOKASC-3. These discrepancies are likely a consequence of our smaller sample size, which may introduce selection effects when compared to the larger APOKASC-3 dataset.

More data in the first dredge-up domain will be an important test of stellar theory. In particular, Li destruction during this phase is a sensitive test of envelope overshooting, which is commonly invoked to explain the mismatch between the observed and predicted red giant branch bump. Existing data is in clear tension with stellar evolution models \citep{2025arXiv250114867C}, but the overlap between seismic masses and Li surveys has been limited. The southern GALAH survey, in particular, did not observed the \textit{Kepler} fields, so we cannot test this directly with our sample. TESS or K2, however, are much more promising in this regard.

We also expect that current and future missions will continue to increase the number of stars in this parameter space that are available for study. In the TESS mission, the full frame image cadence decreased to 10 minutes and 3 minutes during the two extended missions. The PLAnetary Transits and Oscillations of stars \citep[PLATO,][]{2014ExA....38..249R} mission should also allow a large number of detection of solar-like oscillations in the subgiant phase given the yield predictions \citep{2024A&A...683A..78G}.
However, the 1440 day duration of the \kep mission continues to be unmatched, and so we expect these near and Super-Nyquist stars to remain a valuable addition to the literature for years to come.
\section{Data availability}
The data from Tables \ref{supnytable},\ref{nytable} as well as the data described in Table \ref{table_table} (see Appendix \ref{app:table_format}) are only available in electronic form at the CDS via anonymous ftp to \url{cdsarc.u-strasbg.fr (130.79.128.5)} or via \url{http://cdsweb.u-strasbg.fr/cgi-bin/qcat?J/A+A/}.

\begin{acknowledgements}
This paper includes data collected by the \emph{Kepler} mission and obtained from the MAST data archive at the Space Telescope Science Institute (STScI). Funding for the \emph{Kepler} mission is provided by the NASA Science Mission Directorate. STScI is operated by the Association of Universities for Research in Astronomy, Inc., under NASA contract NAS 5–26555. 
R.A.G. acknowledges the support from the GOLF and PLATO Centre National D'{\'{E}}tudes Spatiales grants. 
S.M.\ acknowledges support by the Spanish Ministry of Science and Innovation with the grant number PID2019-107061GB-C66, and through AEI under the Severo Ochoa Centres of Excellence Programme 2020--2023 (CEX2019-000920-S). B.L. thanks \'Ecole Normale Supérieure Paris-Saclay (France) for its stipend that made possible to fund a long stay abroad.
SM, DHG, DGR, and RAG acknowledge support from the Spanish Ministry of Science and Innovation with the grant no. PID2023-149439NB-C41.
PGB acknowledges support by the Spanish Ministry of Science and Innovation with the \textit{Ram{\'o}n\,y\,Cajal} fellowship number RYC-2021-033137-I and the number MRR4032204. 
PGB, DHG, DGR, and RAG acknowledge support from the Spanish Ministry of Science and Innovation with the grant no. PID2023-146453NB-100 (\textit{PLAtoSOnG}).
AS acknowledges support by the Spanish Ministry of Science, Innovation and Universities through the grant PID2023-149918NB-I00 and the program Unidad de Excelencia Mar\'{i}a de Maeztu CEX2020-001058-M, and Generalitat de Catalunya through grant 2021-SGR-1526.
MHP acknowledges support from NASA grant 80NSSC24K0637.  MHP acknowledges support from the Fundaci\'on Occident and the Instituto de Astrof\'isica de Canarias under the Visiting Researcher Programme 2022-2025 agreed between both institutions. DGR acknowledges support from the Juan de la Cierva program under contract JDC2022-049054-I. DHG acknowledges  the support of a fellowship from ”la Caixa” Foundation (ID 100010434). The fellowship code is LCF/BQ/DI23/11990068.

\\
\textit{Software:} AstroPy \citep{astropy:2013,astropy:2018}, Matplotlib \citep{Hunter:2007}, NumPy \citep{harris2020array}, SciPy \citep{2020SciPy-NMeth}

\end{acknowledgements}

\bibliographystyle{aa}
\bibliography{BIBLIO_full}

\newpage

\begin{appendix}
\section{Flowchart}\label{app:flowchart}
\begin{figure}[h]
    \centering
    \includegraphics[scale=0.29]{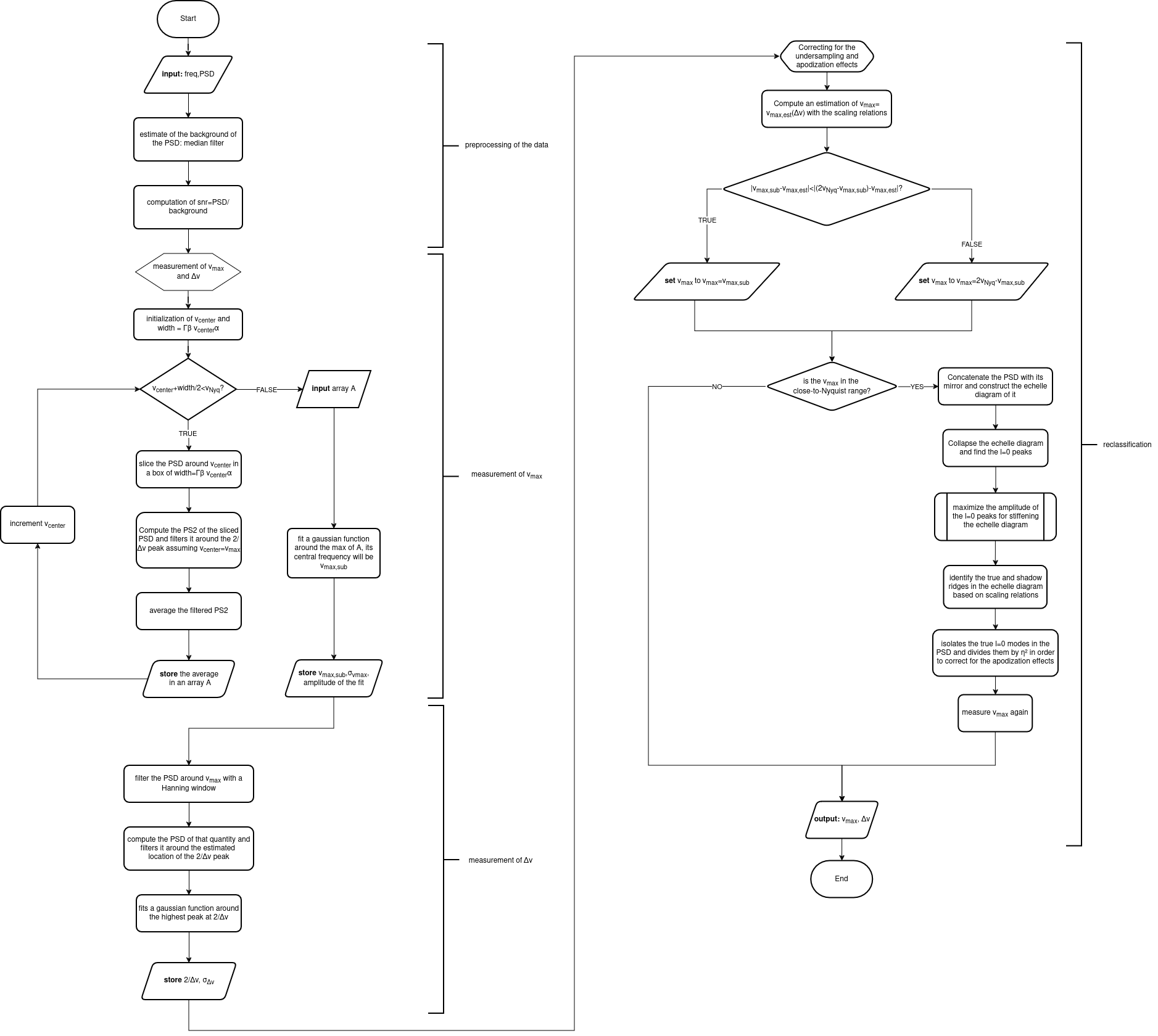}
    
    \caption{Flowchart of the method used.}
    \label{fig:flowchart}
\end{figure}
\newpage ~\newpage
\section{Scaling relation used for this article}\label{app:scalrel}
\begin{figure}[ht]
    \centering
    \includegraphics[width=\linewidth]{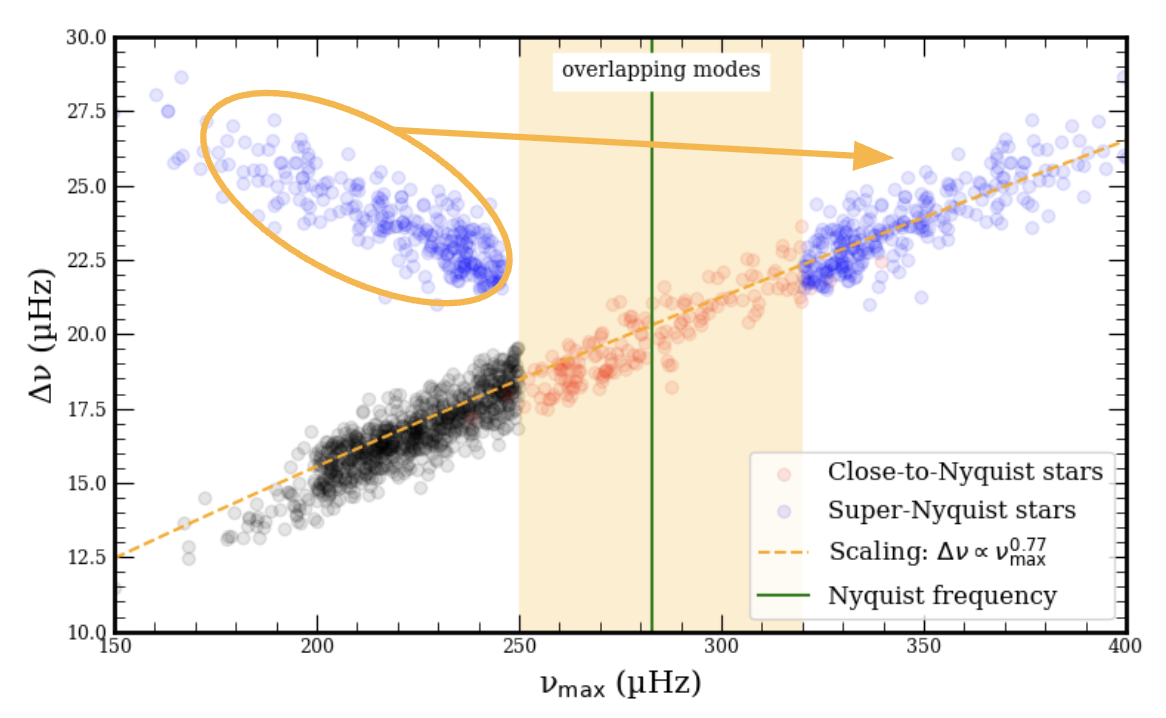}
    \caption{Stars recovered in this work, shown along with the scaling relation adopted in our analysis (orange dashed line). The vertical green line marks the Nyquist frequency. To the left of this line, sub-Nyquist stars are shown in black and super-Nyquist stars, initially misidentified with a frequency $\numax = \nusub$, are shown in blue. To the right of the Nyquist frequency, the same super-Nyquist stars are plotted in blue with their corrected values $\numax = 2\times \nuny-\nusub$. Stars located near the Nyquist frequency, where mode overlap is significant, are shown in red within the yellow shaded region.}
    \label{fig:enter-label}
\end{figure}

\section{Apodization effects}\label{app:apo}
When a point on a lightcurve is measured, the star's luminosity flux is integrated on $T_{\mathrm{samp}}=\frac{1}{\nu_{\mathrm{samp}}}=29.4$ min. Let us introduce the function that represents the light received from the star by the sensor of \textit{Kepler} at a time $t$: $L:t\mapsto L(t)$ as well as the function $V$ that represents the discrete lightcurve given by \textit{Kepler} with a point every sampling period. Is also introduced the function $$\Pi \left({\frac {t}{a}}\right)=\left\{{\begin{array}{rl}0,&{\text{if }}|t|\geq{\frac {a}{2}}\\1,&{\text{if }}|t|<{\frac {a}{2}}\end{array}}\right.\;.$$ Then one point of the lightcurve at $\tau$ corresponds to:
\begin{equation}
\begin{split}&V\left(\tau: t\mapsto t+\frac{T_{\mathrm{samp}}}{2}\right) \propto \int_{t}^{t+T_{\mathrm{samp}}}L(t')\mathrm{d}t' \\&= \int_{-\infty}^{\infty}L(t')\Pi \left({\frac {t'-\tau}{T_{\mathrm{samp}}}}\right)\mathrm{d}t'=L(\tau)*\Pi\left(\frac{\tau}{T_{\mathrm{samp}}}\right)\quad.\end{split}\end{equation} 
The symbol $*$ refers to the convolution product. This means that the lightcurve can be expressed as:
\begin{equation}
    \mathrm{LC}\propto\Sha_{T_{{\mathrm{samp}}}}\left(\tau\right)\times V(\tau)=\Sha_{T_{\mathrm{samp}}}(\tau)\times\left(L(\tau)*\Pi\left(\frac{\tau}{T_{\mathrm{samp}}}\right)\right)\quad,\end{equation}
where $\Sha_T$ is the Dirac comb function of period $T$. Hence the PSD of the lightcurves is:
\begin{equation}\mathrm{PSD}\propto \left|\Sha_{\frac{1}{T_{\mathrm{samp}}}}(\nu)*\mathcal{F}[L](\nu)\right|^2\times \sinc^2(\pi T_{\mathrm{samp}}\nu)\quad.\end{equation}
With $\mathcal{F}[L]$ the Fourier transform of the function $L$. Besides by definition $\nu_{\mathrm{samp}}=2\nuny$. Hence the apodization effect is due to the term $\eta^2=\sinc^2(\pi T_{\mathrm{samp}}\nu)=\sinc^2\left(\frac{\pi\nu}{2\nuny}\right)$.\\
That apodization leads to a deformation of the Gaussian envelope of the modes that can cause biases to the measure of $\numax$ since the modes' amplitudes are no longer symmetrically distributed.

\section{Analysis of TESS data}
\label{app_tess}
\begin{figure}[h!]
    \centering
    \includegraphics[width=\linewidth]{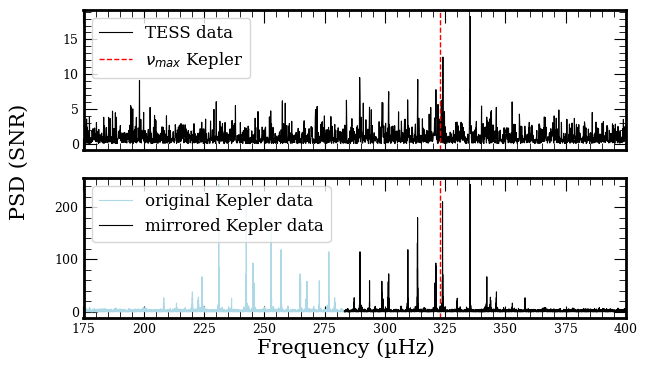}
    \caption{PSD of TIC~123498033/KIC~8279949 as observed by \emph{Kepler} long cadence and TESS.}
    \label{fig:TESSVSKEP}
\end{figure}
We analyzed TESS light curves using the Simple Aperture Photometry (SAP) flux extracted from QLP data. These fluxes were corrected with \texttt{PyTADACS}, an adaptation of the \emph{Kepler} Asteroseismic Data Calibration Software \citep[\texttt{KADACS};][]{2011MNRAS.414L...6G} specifically tailored for TESS observations. \texttt{PyTADACS}, currently under development, provides enhanced calibration for TESS data \citep[see][for further details]{2024tsc3.confE..32G,2024tkas.confE.123G}. From the 2 104 \emph{Kepler} targets analyzed in this work, we only found 1 140 available in TESS QLP data up to sector 80. After applying PyA2Z, only 9 stars had detectable oscillations. Results are reported in Table~\ref{Tab:Tess_results}.
\onecolumn
\begin{table}[htbp]
\centering
\caption{\label{Tab:Tess_results} Comparison of TESS and \textit{Kepler} retrieved global seismic parameters.}
\begin{tabular}{ccSSSSS}
\hline\hline
KIC & TIC & {$\nu_{\mathrm{max,TESS}}$} & {$\nu_{\mathrm{max,}\textit{Kepler}}$} & {$\Delta \nu_{\mathrm{TESS}}$} & {$\Delta \nu_{\textit{Kepler}}$} & {\# TESS} \\
& & {(µHz)} & {(µHz)} & {(µHz)} & {(µHz)} & {Sectors}\\
\hline
3936801  & 120572957 & 225 \pm 8  & 225 \pm 9  & 16.7 \pm 0.8 & 17.0 \pm 0.2 & 5\\
4351319  & 121215521 & 380 \pm 12 & 375 \pm 12 & 24.9 \pm 0.9 & 24.4 \pm 0.3 & 7\\
4913049  & 121659257 & 219 \pm 9  & 222 \pm 9  & 17.5 \pm 1.1 & 17.0 \pm 0.2 & 7\\
7450230  & 270525099 & 212 \pm 8  & 219 \pm 8  & 16.6 \pm 1.2 & 16.7 \pm 0.8 & 6\\
7741472  & 158423349 & 203 \pm 8  & 205 \pm 8  & 16.3 \pm 1.4 & 16.2 \pm 0.9 & 8\\
7812552  & 158720264 & 211 \pm 9  & 214 \pm 9  & 18.6 \pm 1.6  & 17.0 \pm 1.0 & 8\\
8279949  & 123498033 & 319 \pm 11 & 323 \pm 11 & 21.9 \pm 1.1 & 22.3 \pm 1.0 & 7\\
10656270 & 158562300 & 238 \pm 10 & 248 \pm 10 & 20.8 \pm 1.2 & 19.2 \pm 0.5 & 8\\
12506768 & 299217712 & 189 \pm 7  & 193 \pm 7  & 14.8 \pm 0.8 & 14.7 \pm 0.7 & 11\\
\hline
\end{tabular}
\end{table}

\section{Columns of the table published with this article}
\begin{table}[htbp]
    \centering
    \caption{Columns included in the final table.}\label{table_table}
    \begin{tabular}{c c}
    \hline\hline
         Label & Contents  \\
          
 \hline
 KIC & Identifier in the \kep\ Input Catalog \\
Cat & Category (close-to-Nyquist, sub-Nyquist, super-Nyquist) \\
SpecSource & Spectroscopic source \\
NQuar & Quarters of \kep\ data \\
Numax, SNumax & $\numax$\ ($\mu$Hz) and $\sigma$ \\
Dnu, SDnu & $\dnu$\ ($\mu$Hz) and $\sigma$ \\
FDnu, SFDnu & Mosser $\fdnu$\ and $\sigma$ \\
Mass, SMass &  Mass ($\msun$) and $\sigma$ \\
Radius, SRadius &  Radius ($\rsun$) and $\sigma$ \\
Loggseis, Sloggseis &  Seismic $\logg$ (cgs) and $\sigma$ \\
Teff, STeff & $\teff$ and $\sigma$ (K)\\
L, SL & Luminosity and $\sigma$ (K)\\
Loggspec, Sloggspec & Spectroscopic $\logg$ (cgs) and $\sigma$ \\
FeH, SFeH & \feh ([M/H]) and $\sigma$ \\
AlpFe, SAlpFe & \afe and $\sigma$ \\
CFe, SCFe & [C/Fe] and $\sigma$ \\
NFe, SNFe & [N/Fe] and $\sigma$ \\
InvRGaia, SInvRGaia & MIST K $\frac{1}{R_{\gaia}}$ and $\sigma$ \\
AgeRGB & Garstec Age (Gyr), RGB \\
SAgeRGB+,SageRGB- & $\pm $Garstec Age $\sigma$ (Gyr), RGB \\
GaiaDR3 & Identifier in the 
\gaia\ Catalog (DR3)\\
TIC & Identifier in the 
TIC Catalog \\
 2MASS & Identifier in the 2MASS Input Catalog \\
 \hline
 \end{tabular}
 \end{table}
Table \ref{table_table} details the columns of the master table that will be available at the CDS. This includes the columns appearing in tables \ref{supnytable},\ref{nytable} and the informations given in table \ref{tab:res}.
\label{app:table_format}
\end{appendix}
\end{document}